\documentclass[bibyear]{aa} 

\usepackage{txfonts}
\usepackage{graphicx}   
\usepackage{amsmath}    
\usepackage[utf8]{inputenc} 
\usepackage{array} 
\usepackage{multicol,rotating} 
\usepackage{longtable} 
\usepackage[section]{placeins}
\usepackage{lscape}

\begin{document}

   \title{Phase reddening on asteroid Bennu from visible and near-infrared spectroscopy}


   \author{S. Fornasier \inst{1,2} \and
          P. H. Hasselmann \inst{1} \and
          J. D. P Deshapriya \inst{1} \and
          M. A. Barucci  \inst{1} \and
          B. E. Clark \inst{3} \and
          A. Praet \inst{1} \and
          V. E. Hamilton  \inst{4} \and
          A. Simon \inst{5} \and
          J-Y. Li\inst{6} \and
          E. A. Cloutis\inst{7} \and
          F. Merlin \inst{1} \and
          X-D. Zou \inst{6} \and
          D. S. Lauretta \inst{8}
          }

   \institute{LESIA, Observatoire de Paris, Universit\'e PSL, CNRS, Universit\'e de Paris, Sorbonne Universit\'e, 5 place Jules Janssen, 92195 Meudon, France\\
              \email{sonia.fornasier@obspm.fr}
         \and
           Institut Universitaire de France (IUF), 1 rue Descartes, 75231 Paris Cedex 05  
          \and  Department of Physics and Astronomy, Ithaca College, Ithaca, NY, USA  
          \and  Southwest Research Institute, Boulder, CO, USA  
          \and Solar System Exploration Division, NASA Goddard Space Flight Center, Greenbelt MD USA
          \and Planetary Science Institute, Tucson, AZ, USA  
          \and Department of Geography, University of Winnipeg, Winnipeg, MB, Canada
          \and Lunar and Planetary Laboratory, University of Arizona, Tucson, AZ, USA 
             }

   \date{Received on September 2020}

 
  \abstract
  {The NASA mission OSIRIS-REx (Origins, Spectral Interpretation, Resource Identification, and Security--Regolith Explorer) has been observing near-Earth asteroid (101955) Bennu in close proximity since December 2018. The spacecraft has collected, on October 2020, a sample of surface material from Bennu to return to Earth.} 
   {In this work, we investigate spectral phase reddening---that is, the variation of spectral slope with phase angle---on Bennu using spectra acquired by the OSIRIS-REx Visible and InfraRed Spectrometer (OVIRS) covering a phase angle range of 8--130$^{o}$. We investigate this process at the global scale and for some localized regions of interest (ROIs), including boulders, craters, and the designated sample collection sites of the OSIRIS-REx mission.}
   {Spectra were wavelength- and flux-calibrated, then corrected for the out-of-band contribution and thermal emission, resampled, and finally converted into radiance factor per standard OVIRS processing. Spectral slopes were computed in multiple wavelength ranges from spectra normalized at 0.55 $\mu$m. 
}
   {Bennu has a globally negative spectra slope, which is typical of B-type asteroids. The spectral slope gently increases in a linear way up to a phase angle of  90$^{\circ}$, where it approaches zero. The spectral phase reddening is monotonic and wavelength-dependent with highest values in the visible range. Its coefficient is 0.00044 $\mu$m$^{-1} ~deg^{-1}$ in the 0.55--2.5 $\mu$m range. \\
For observations of Bennu acquired at high phase angle (130$^{\circ}$), phase reddening increases exponentially, and the spectral slope becomes positive. Similar behavior was reported in the literature for the carbonaceous chondrite Mukundpura in spectra acquired at extreme geometries. Some ROIs, including the sample collection site, Nightingale, have a steeper phase reddening coefficient than the global average, potentially indicating a surface covered by fine material with high micro-roughness.}
   {The gentle spectral  phase reddening effect on Bennu is similar to that observed in ground-based measurements of other B-type asteroids, but much lower than that observed for other low-albedo bodies such as Ceres or comet 67P/Churyumov-Gerasimenko. Monotonic reddening may be associated with the presence of fine particles at micron scales and/or of particles with fractal structure that introduce  micro- and sub-micro roughness across the surface of Bennu.}
   {}

   \maketitle
%

\section{Introduction}

OSIRIS-REx (Origins, Spectral Interpretation, Resource Identification, and Security--Regolith Explorer) is a NASA sample return mission devoted to the study of the primitive near-Earth asteroid (101955) Bennu. Launched in September 2016, the mission arrived at the target in December 2018 and, since then, has been examining Bennu from different angles and altitudes to globally characterize its physical, chemical, and geological properties and to find a suitable site for sample collection. Higher spatial resolution observations have also been carried out to characterize in detail four candidate sample collection sites (called Sandpiper, Osprey, Kingfisher, and  Nightingale).  The primary and back-up sampling sites, Nightingale and Osprey, respectively, were selected in December 2019. The spacecraft collected a sample from Nightingale in October 2020,

The spacecraft carries a suite of instruments including scientific imaging cameras (OCAMS, the OSIRIS-REx Camera Suite), spectrometers (OVIRS, the OSIRIS-REx Visible and InfraRed Spectrometer, 0.4--4.3 $\mu$m; OTES, the OSIRIS-REx Thermal Emission Spectrometer, 5.5--100 $\mu$m; and REXIS, the Regolith X-Ray Imaging Spectrometer), and a scanning lidar system (OLA, the OSIRIS-REx Laser Altimeter), in addition to the sampling device (Lauretta et al. 2017).

 \begin{table}
      \caption[]{Observing conditions for the OVIRS data analyzed in this paper. The heading "Obs." indicates the mission phase, in which PS is Preliminary Survey, BBD is Baseball Diamond, EQ is Equatorial Station, and Recon is Reconnaissance. We excluded from our analysis the data acquired during the BBD1 \& 2 observations on 2019 March 7$^{th}$ and 14$^{th}$, respectively, because they are affected by saturation or calibration issues. The BBD2 was re-flown in September, and those data are included. The quantity $\Delta$ is the spacecraft-Bennu distance, {\it Res.} the spatial resolution, $\alpha$ the phase angle. The heading {\it Nspec} indicates the number of spectra used in this study from each dataset satisfying the criteria on fill factor, incidence and emission angles, and radiance detailed in section 2.}
         \label{observations}
{\small
\centering                       
\begin{tabular}{l c c c c c}       
\hline\hline                
Obs. & Date & $\Delta$  & Res. & $\alpha$  & Nspec \\
     &      & (km)          & (m)      & [$^o$] & \\ \hline
PS &  9 Dec. 2018 & 11.3 & 45 & 91.3-92.1 &  423 \\ 
PS &  12 Dec. 2018 & 10.0 & 40 & 38.3-51.9  & 8946 \\
PS &  13 Dec. 2018 & 9.9 & 40 & 38.3-51.8  & 1741 \\
PS &  16 Dec. 2018 & 11.3 & 45 & 90.4-91.3  & 237 \\
PS &  17 Dec. 2018 & 11.0 & 44 & 89.7-91.3  & 655 \\
BBD3 & 21 March 2019 & 3.6 & 14 & 30.1-31.3 & 16207 \\
EQ1 & 25 April 2019 & 4.9 & 20 & 42.6-48.0  & 7431 \\
EQ2 & 2-3 May 2019 & 4.9 & 20 & 128.5-132.6    & 641 \\ 
EQ3 & 9 May 2019 & 4.8 & 19 & 7.8-10.4     & 7109 \\ 
    & 11 May 2019 & 5.3 & 21 & 50.9-62.6     & 4688 \\ 
EQ4 & 16 May 2019 & 4.8 & 19 & 29.7-30.4     & 6040 \\ 
EQ5 & 23 May 2019 & 4.8 & 19 & 90.2-92.6    & 907 \\ 
    & 26 May 2019 & 4.3 & 17 & 76.0-88.3    & 3391 \\
EQ6 & 30-31 May 2019 & 5.1 & 20 & 128.0-132.7    & 480 \\ 
EQ7 & 6 June 2019 & 4.9 & 20 & 90.1-93.3     & 445 \\
BBD2 & 26 Sept. 2019 &  3.8 & 15 & 7.3-10.8  & 8209 \\
re-fly &            &       &   &               &      \\
Recon   &  5 Oct. 2019 & 1.35  & 5.4 & 29.4-50.6  & 3084 \\
Recon    & 12 Oct. 2019 & 1.22 & 4.9 & 27.1-44.2  & 4711 \\
Recon    & 19 Oct. 2019 & 1.29  & 5.2 & 37.0-47.0  & 4705 \\
Recon    & 26 Oct. 2019 & 1.35 & 5.4 & 21.5-51.8  & 3226 \\
Recon    & 21-22 Jan. 2020 & 0.87 & 3.5 & 65.8-88.2  & 5585 \\
Recon    & 11 Feb. 2020    & 0.96 & 3.8 & 12.7-59.1  & 9376 \\
\hline                                   
\end{tabular}
}
   \end{table}

Extensive ground-based studies of asteroid Bennu in support of the mission have revealed a roughly spherical body with an equatorial bulge (Lauretta et al. 2015). 
The top-like shape has been confirmed by OCAMS images, which also revealed an unexpectedly rugged surface (Lauretta et al. 2019a; Barnouin et al. 2019). Bennu has a very dark surface with a global average geometric albedo of 4.4$\pm$0.5\% (DellaGiustina et al. 2019, Hergenrother et al. 2019), but showing local variations up to $\sim$ 26\% in some pyroxene-rich rocks a few square meters in size, which have exogenous origin (DellaGiustina et al. 2020a). Repeated episodes of particle ejections have been observed, indicating that Bennu is an active asteroid (Lauretta et al. 2019b).  \\
The OSIRIS-REx data have confirmed that Bennu is a rubble-pile asteroid that reaccumulated from the catastrophic disruption of a primordial parent body (Walsh et al. 2019; Barnouin et al. 2019). The composition of this asteroid includes hydrated minerals, carbon species (organics, carbonates), and magnetite (Lauretta et al. 2019a; Hamilton et al. 2019, Simon et al. 2020a). Visible (VIS)\ to near-infrared (NIR) and emissivity spectra are similar to those of aqueously altered CM- or possibly CI-type carbonaceous chondrites (Hamilton et al. 2019). \\

In this paper, we investigate the phase reddening effect (i.e., the variation of the spectral slope with phase angle) from OVIRS spectra of the global Bennu surface and of some local regions of interest (ROIs), including boulders, craters, and the four candidate sampling sites. \\
Phase reddening is a common behavior observed on many Solar System objects, including asteroids (Taylor et al. 1971; Clark et al. 2002; Magrin et al. 2012; Ciarniello et al. 2017; Longobardo et al. 2018; Li et al. 2019; Sanchez et al. 2012, Perna et al. 2018; Lantz et al. 2017; Li et al. 2015), comets (Fornasier et al. 2015; Ciarniello et al. 2015; Longobardo et al. 2017), the Moon (Gehrels et al. 1964), Mercury (Warell \& Bergfors, 2008), and planetary satellites (Nelson et al. 1987; Cuzzi et al. 2002, Filacchione et al. 2012). This phenomenon has been attributed to multiple scattering at intermediate to high phase angles and/or small-scale surface roughness, thus its investigation helps to constrain the physical properties, in terms of particle size and roughness, of the surface of Bennu.

\section{Observations and methodology}

%
%
 \begin{figure*}
   \centering
  \includegraphics[width=0.99\textwidth,angle=0]{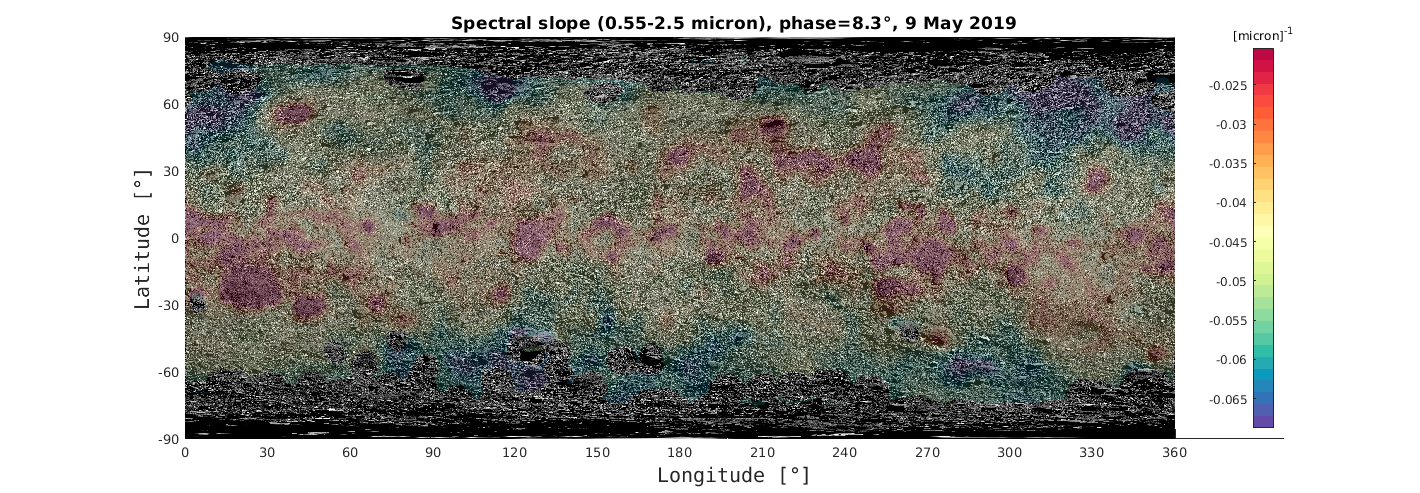}
\includegraphics[width=0.99\textwidth,angle=0]{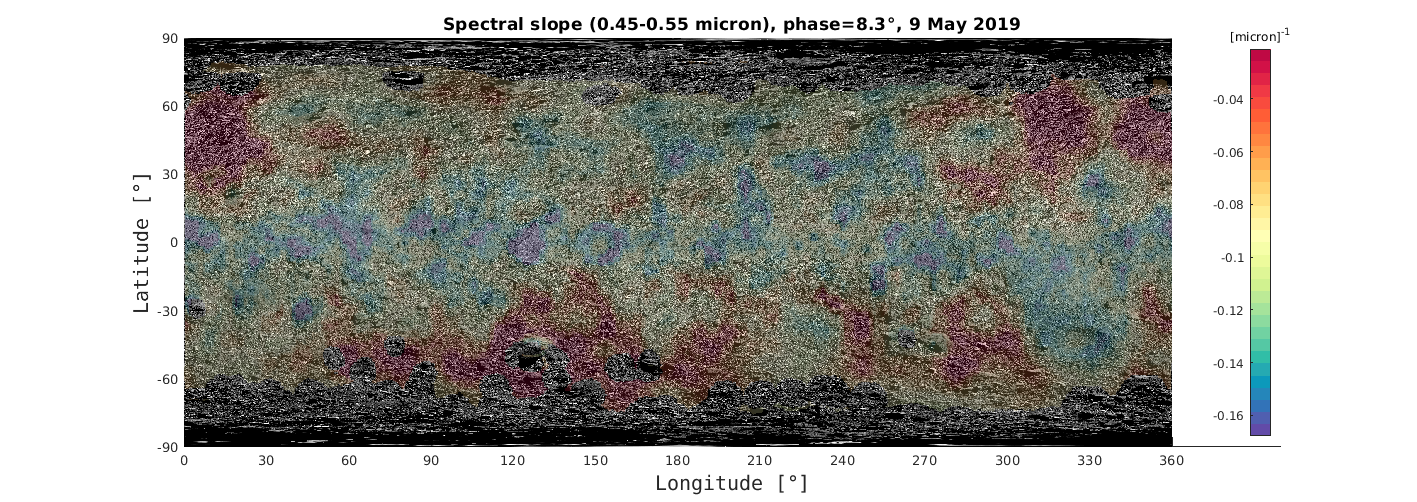}
   \caption{Maps of the spectral slope, evaluated in the 0.55--2.5 $\mu$m and 0.45--0.55 $\mu$m ranges, for the observations acquired during EQ3 (9 May 2019). Maps acquired at different phase angles have a similar behavior in relative terms to those shown here.}
              \label{maps2a}
    \end{figure*}
The OVIRS instrument has five filter segments and a  field of view of 4 mrad (Reuter et al. 2018). We used Level 3 spectra processed by the OVIRS calibration pipeline (Reuter et al. 2019) at the OSIRIS-REx Science Processing and Operations Center. These data include the absolute calibration in physical units (W cm$^{-2}$ sr$^{-1}$ $\mu$m$^{-1}$); the removal of the out-of-band contribution; the resampling of the wavelength axis to a spectral resolution of 2 nm and 5 nm for the ranges 0.4 to 2.4 $\mu$m and 2.4 to 4.3 $\mu$m, respectively; and the thermal emission correction by fitting the thermal tail with a blackbody curve. The data are ultimately converted into radiance factor $I/F$, where I is the observed scattered radiance, and $\pi \times F$ is the incoming  solar irradiance at the heliocentric distance of Bennu. The illumination and geometric conditions for each  spectrum were derived from the intersections of the OVIRS field of view with the stereophotoclinometric shape model of Bennu (v20 for the global analysis, and v22 for the Reconnaissance observation phase data; Barnouin et al. 2019; 2020). The OVIRS performance and calibration are reported in Simon et al. (2018).

\subsection{Global analysis}

To investigate the global spectral properties of Bennu, we analyzed OVIRS data (Reuter et al. 2019) obtained between December 2018 and September 2019, covering a phase angle range of 8--130$^o$ at a spatial resolution varying from 40 m/pixel to $\sim$ 16 m/pixel (Table~\ref{observations}). These observations were acquired during the OSIRIS-REx mission phases Preliminary Survey (PS, December 2018) and Detailed Survey, the latter of which included the subphases Baseball Diamond (BBD; March and September 2019) and Equatorial Stations (EQ; seven stations at different local solar times between April 25 and June 6, 2019) (Lauretta et al. 2017). 

Each observing run reported in Table~\ref{observations} typically encompassed several thousands of spectra. We selected those satisfying the following conditions: (a) the fill factor should be 100\%; that is, the OVIRS spot should be entirely on the surface of Bennu; (b) the incidence and emission angles should both be  $<$ 75$^{o}$ to avoid extreme illumination conditions; and (c) the radiance factor should be $>$ 0.001 to avoid spectra with low signal-to-noise ratios for observations acquired at high phase angles, and to limit the occurrence of pixels in projected shadows.

We normalized each spectrum at 0.55 $\mu$m, and then we computed the spectral slope in multiple wavelength ranges as follows:
\begin{equation}
 Slope = \frac{Refl ({\lambda_2}) - Refl ({\lambda_1}) }{ Refl(0.55) \times (\lambda_2 - \lambda_1)}, \end{equation}

where $\lambda$ is the wavelength and the spectral ranges considered are 0.45--0.55 $\mu$m,  0.55--0.86 $\mu$m, 0.55--2.5 $\mu$m, 1.08--1.70 $\mu$m, and 1.7--2.5 $\mu$m.

\subsection{Analysis of regions of interest}

We also investigated the phase reddening and linear phase function on localized ROIs including craters, outcrops, and boulders having peculiar morphology and/or outstanding reflectance or spectral properties compared to the global properties of Bennu.  We studied 40 ROIs that have satisfactory spectral coverage in terms of OVIRS spatial footprint and phase angle range. For each ROI, data were selected from the OVIRS database covering the March 
2019--February 2020 period, thus including both medium and high spatial resolution observations; these last were acquired mostly during the characterization of the four candidate sampling sites (Table~\ref{observations}). \\
We verified the OVIRS footprint locations for every ROI by projecting them onto the OCAMS global image mosaic (Bennett et al. 2020) using the 
J-Asteroid software (Christensen et al. 2018). We thus checked whether the conditions
of observation were suitable and whether the spectral data were well placed over the selected features.  We adopted the same filtering criteria for incidence and emission angles and minimal reflectance as done for the Bennu global analysis.  As very few ROIs have been observed at phase angle ($\alpha$) $>90^{\circ}$, we removed these high phase angle observations to obtain a homogeneous analysis of the different ROIs.  \\
Finally, we applied a $3\sigma$ filter with respect to the mean spectral slope of the features to remove extreme outliers. The spectral slopes and their respective uncertainties were computed for the various spectral
ranges previously described for the global analysis. However, owing to the 0.65 $\mu$m segment jump between two OVIRS filters that affects some high spatial resolution data even after the $3\sigma$ filter selection, we estimated the "visible range" phase coefficient in the 0.7--1.06 $\mu$m range for the ROIs, rather than in the 0.55-0.86 $\mu$m range as used for the global Bennu analysis.    \\

\subsubsection{Phase function slope} 

For the ROIs, we also looked for possible correlations between the slope reddening coefficients and the phase function slope, as both quantities are related to multi-scale surface roughness (Schr\"oder et al. 2014; Shepard \& Helfenstein, 2011). 
The phase function adopted in this work is a linear model of the equigonal albedo, that is, the reflectance after correction by the Akimov disk function. The linear coefficient estimates the phase function slope, while the y-intercept is
the linear albedo, that is, the estimated normal albedo at zero phase, excluding the contribution of the exponential-like surge in reflectance at small phase angles related to the opposition effect.
 We adopted a simple linear model for the phase function, considering that most of the data from the ROIs have a limited phase angle coverage; in particular, no data were taken at phase angles $<$ 8$^{\circ}$ covering the opposition surge. We also considered that in complex radiative transfer models some parameters are correlated and difficult to disentangle (Schmidt \& Fernando,
2015; Li et al. 2015). 

Before tracing the phase function slope, we corrected all
OVIRS data with the Akimov disk function to subtract the topographic-photometric
contribution from the radiance factor. This function depends on the
($i$, $e$, $\alpha$) scattering angles and has been mathematically
derived for quasi-fractal self-affine rough surfaces. The Akimov function
has been shown to provide satisfactory disk correction for atmosphereless
planetary surfaces (Shkuratov et al. 2011; Schr\"oder et al. 2017),
including that of Bennu (Zou et al. 2020). Zou et al. (2020) 
provide a detailed investigation of different phase function models for Bennu from OVIRS observations.

The Akimov disk function is given by

\begin{equation}
D_{(\alpha,\beta,\phi)}=\cos\frac{\alpha}{2}\cos\left(\frac{\pi}{\pi-\alpha}\left(\phi-\frac{\alpha}{2}\right)\right)\frac{\cos^{\alpha/(\pi-\alpha)}\beta}{\cos\phi}
,\end{equation}

where $\alpha$ is the phase angle, and $\beta_{(i,e)}$ and $\phi_{(i,e)}$ are the photometric latitude
and longitude, respectively (Shkuratov et al. 2011). 

We studied the albedo and slope of the phase function at two wavelengths, $0.55\pm0.05\ \mu m$ and $1.55\pm0.05\ \mu m$.
Both parameters were estimated using the orthogonal distance linear
regression\footnote{\url{https://docs.scipy.org/doc/scipy/reference/odr.html}}
(ODLR; Boggs and Rogers, 1990), a technique that  minimizes the
least-squares fit between data and model.  Phase function parameters
are listed in the Table~\ref{phasefunction} of appendix A.\\

\subsection{Phase reddening and spectral slope at $0^{\circ}$ phase angle}

A linear model was applied to describe the spectral slope dependence on the phase angle.
From the spectral slopes computed at different phase
angles, we calculated the phase reddening coefficient and the spectral slope at
$0^{\circ}$ phase angle. These values are given by the angular coefficient and the intercept of the linear model, both for the global Bennu and local ROIs investigation via the ODLR technique. In our analysis we only considered the ROIs that have data covering at least the 8--50$^{\circ}$ phase angle range, given that undersampled data can lead to erroneous phase reddening parameters. The errors in the phase reddening parameters were estimated from the spectral slope errors by the ODLR technique.


 \begin{figure}
   \centering
  \includegraphics[width=0.44\textwidth,angle=0]{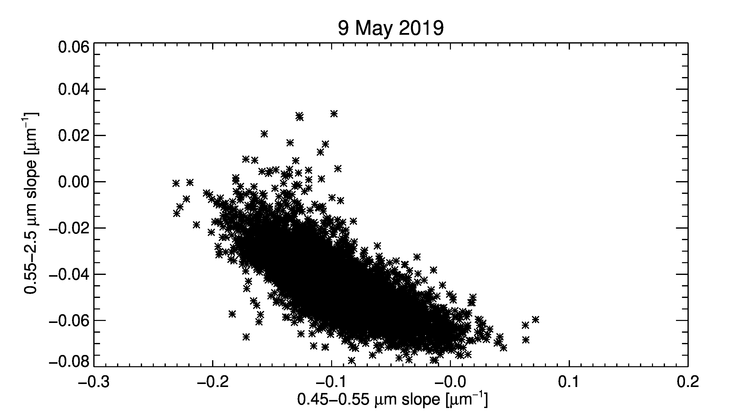}
  \includegraphics[width=0.44\textwidth,angle=0]{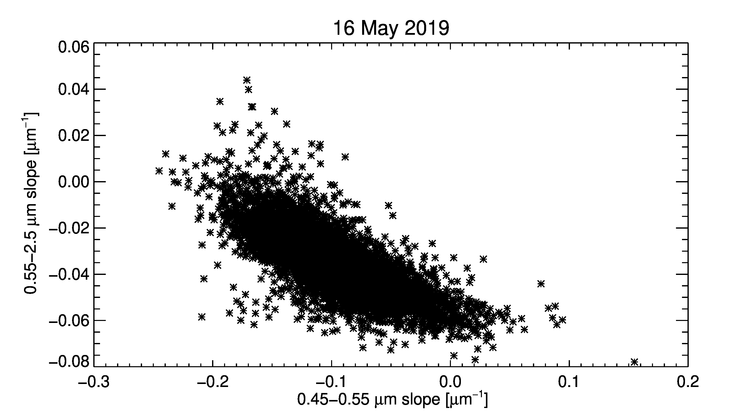}
   \caption{Distribution of the 0.55--2.5 $\mu$m spectral slope vs. that of 0.45--0.55 $\mu$m spectral slope for the global observations. These observations were acquired at EQ3 (top) and EQ4 (bottom) at phase angles of $\sim$ 8$^{\circ}$ and 30$^{\circ}$, respectively, showing the common increase of the two spectral slopes with phase angle, while remaining strongly anticorrelated with one another. A similar correlation between the two slopes is observed for the other datasets.}
              \label{correlation}%
    \end{figure}
 \section{Bennu spectral behavior}

  \begin{figure*}
   \centering
  \includegraphics[width=0.95\textwidth,angle=0]{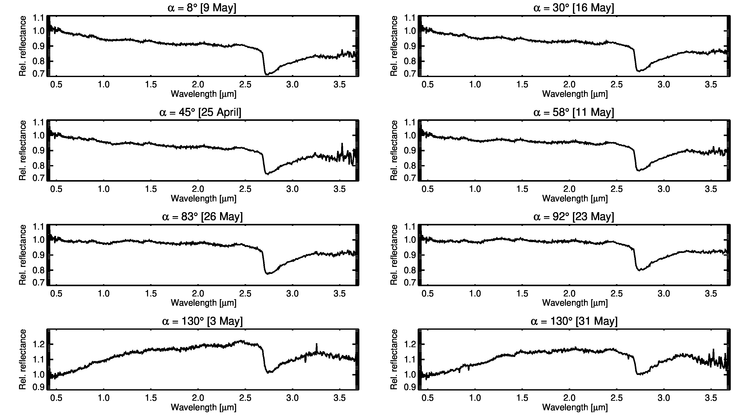}
   \caption{Averaged Bennu reflectance spectra, normalized at 0.55 $\mu$m, for different observing geometries from April--May 2019 Equatorial Station data.}
              \label{spectra}
    \end{figure*}

 \begin{figure}
   \centering
  \includegraphics[width=0.45\textwidth,angle=0]{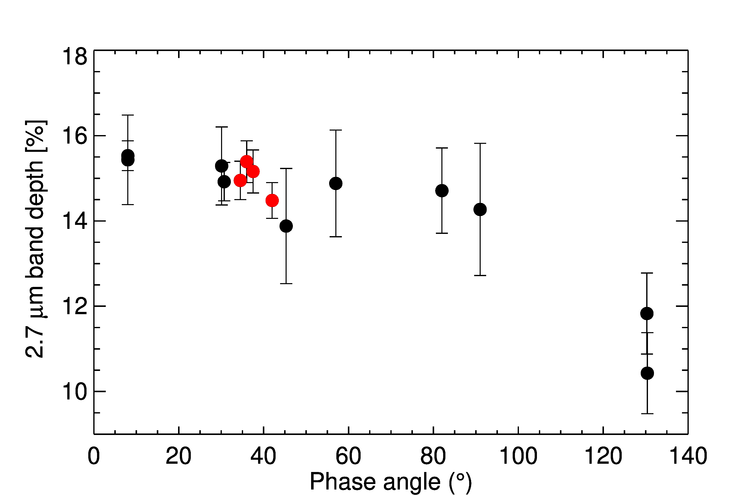}
   \caption{Variation of the 2.74 $\mu$m band depth with the phase angle. The values are the median and standard deviation of the 2.74 $\mu$m band depth for the different observing runs (Table~\ref{observations}). The black points indicate the global observations acquired between March and September 2019, while the red points indicate the higher-spatial-resolution observations of the four candidate sampling sites from October 2019 data.}
              \label{band2e7}%
    \end{figure}

  \begin{figure*}
   \centering
  \includegraphics[width=0.9\textwidth,angle=0]{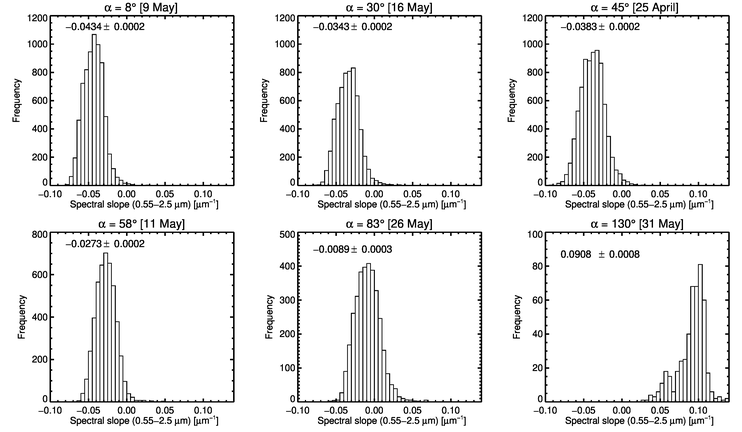}
   \caption{Histograms of the 0.55-2.5 $\mu$m spectral slope for different observing geometries from April--May 2019 Equatorial Station data. The mean values and the associated errors, defined as the standard deviation divided by $\sqrt(n)$, where {\it n} is the number of spectra for a given observing run satisfying the selection criteria mentioned in section 2, are reported for each dataset.}
              \label{histo}%
    \end{figure*}

Bennu was taxonomically classified as a B-type asteroid from ground-based observations (Clark et al. 2011). This classification has been confirmed by the OVIRS and OCAMS data, which show an overall negative slope at low-moderate phase angles typical of the B type (Hamilton et al. 2019; DellaGiustina et al. 2019, 2020b; Simon et al. 2020a). \\
Our analysis confirms that the average  Bennu 0.55--2.5 $\mu$m spectral slope is negative; we find --0.04343 $\mu$m$^{-1}$ at phase $\sim$ 8$^{\circ}$. According to the de Leon et al. (2012) classification of B-type asteroids, Bennu has a spectral behavior intermediate between the G4 and G5 subgroups.  We show in Fig.~\ref{maps2a} the map of the Bennu spectral slope values in the 0.55--2.5 $\mu$m and 0.45--0.55 $\mu$m wavelength ranges from the observations acquired during EQ3 at phase angle 8$^{o}$. We chose this dataset because it is one of the best in terms of Bennu surface coverage and high signal-to-noise ratio.

We observe local spectral variations on the surface of Bennu (Fig.~\ref{maps2a}), as reported in previous work (Simon et al. 2020a; DellaGiustina et al. 2020a, 2020b; Barucci et al. 2020).
If we consider the 0.55--2.5 $\mu$m range, the bluer regions (those having more negative slopes) are found at high latitudes, whereas redder regions (less negative slopes) are present mostly at low latitudes, in the equatorial region.  The reddest regions are associated with specific structures, for example: a $\sim$ 95-m boulder named Roc Saxum (lat = -24$^o$, lon = 28$^o$), which is the reddest feature and one of the darkest on the surface of Bennu (DellaGiustina et al. 2020b), and craters such as the sampling site Nightingale (lat = 55.4$^o$, lon = 42.3$^o$). Similar latitudinal patterns in the spectral slope distribution across the surface of Bennu have been identified by Barucci et al. (2020) from a multivariate statistical analysis of OVIRS data, and by Li et al. (2020) from photometric modeling of OVIRS data. 

We also find an anticorrelation  between the spectral slopes evaluated in the 0.45--0.55 $\mu$m ({\it S1}) and  0.55--2.50 $\mu$m ({\it S2}) wavelength ranges (Figs.~\ref{maps2a} and ~\ref{correlation}). This anticorrelation is reported also by Li et al. (2020) and Simon et al. (2020b) and attributed to the presence of a $\sim$ 0.55 $\mu$m band consistent with iron oxides, which is deepest in the reddest areas (Simon et al. 2020b). \\
To quantify this anticorrelation, we calculated the Spearman rank correlation (Spearman,  1904). This calculation gives a two-element vector containing the rank correlation coefficient ($R$) and the  two-sided significance of its deviation from zero ($p-$value). Strong correlations should have $p-$value $<$ 0.01  and $|R| >$ 0.6. In the case of $S1$ versus $S2$, we find a strong anticorrelation with $R$ = --0.79 and $p-$value $< 10^{-5}$ (Fig.~\ref{correlation}) for both the EQ3 and EQ4 observations. A similar anticorrelation is also observed in the other datasets acquired at different phase angles. Conversely, we did not find any correlation between the spectral slopes and the reflectance.\\

 \begin{table*}
      \caption[]{Averaged spectral slopes and the associated errors for the different observing runs, evaluated in five different ranges: S1 in the 0.45--0.55 $\mu$m range, S2 in the 0.55--2.50 $\mu$m range, S3 in the 0.55--0.86 $\mu$m range, S4 in the  1.08--1.70 $\mu$m range, and S5 in the  1.7--2.5 $\mu$m range.}
         \label{meanslope}
\centering      
{\footnotesize
\begin{tabular}{l c c c c c c c c c c c}       
\hline\hline                
Day            & $\alpha$        &   S1     &   errS1  &   S2     &   errS2  &   S3     &   errS3  &   S4     &   errS4 &   S5     &   errS5\\
              & [$^{\circ}$]     & [$\mu$m$^{-1}$]& [$\mu$m$^{-1}$]& [$\mu$m$^{-1}$]& [$\mu$m$^{-1}$]& [$\mu$m$^{-1}$]& [$\mu$m$^{-1}$]  & [$\mu$m$^{-1}$]& [$\mu$m$^{-1}$] & [$\mu$m$^{-1}$]& [$\mu$m$^{-1}$]\\ \hline
      9 Dec. 2018 &   91.75 &     -0.0040 &   0.0025 &  -0.0174 &   0.0007 &  -0.0368 &   0.0042 &  -0.0077 &   0.0007 &  -0.0091 &   0.0007\\ 
     12  Dec. 2018 &   46.57 &     -0.0988 &   0.0030 &  -0.0372 &   0.0004 &  -0.0959 &   0.0009 &  -0.0285 &   0.0005 &   0.0169 &   0.0004\\ 
     13 Dec. 2018 &   44.64 &     -0.0273 &   0.0010 &  -0.0428 &   0.0003 &  -0.0957 &   0.0010 &  -0.0386 &   0.0004 &   0.0471 &   0.0004\\ 
     16 Dec. 2018 &   90.83 &     -0.0187 &   0.0023 &  -0.0159 &   0.0008 &  -0.0680 &   0.0058 &  -0.0080 &   0.0009 &  -0.0073 &   0.0008\\ 
   21 March 2019 &   30.38 &     -0.0665 &   0.0005 &  -0.0408 &   0.0001 &  -0.0876 &   0.0004 &  -0.0204 &   0.0001 &  -0.0078 &   0.0001\\
    25 Apr. 2019 &   45.00 &      -0.0630 &   0.0008 &  -0.0385 &   0.0002 &  -0.0789 &   0.0014 &  -0.0113 &   0.0002 &  -0.0169 &   0.0002\\ 
      3 May 2019 &  130.61 &      0.1574 &   0.073 &   0.0902 &   0.0030 &   0.2118 &   0.0075 &   0.0392 &   0.0020 &   0.0411 &   0.0010\\ 
     9 May 2019 &    8.25 &      -0.0980 &   0.0005 &  -0.0434 &   0.0002 &  -0.1114 &   0.0004 &  -0.0209 &   0.0002 &  -0.0149 &   0.0002\\ 
    11 May 2019 &   58.28 &      -0.0695 &   0.0013 &  -0.02792 &   0.0002 &  -0.05175 &   0.0015 &  -0.0094 &   0.0004 &  -0.0171 &   0.0003\\ 
     16 May 2019 &   29.88 &     -0.0986 &   0.0006 &  -0.0343 &   0.0002 &  -0.0866 &   0.0007 &  -0.0165 &   0.0002 &  -0.0150 &   0.0002\\
     23 May 2019 &   91.72 &     -0.0091 &   0.0035 &  -0.0118 &   0.0080 &  -0.0168 &   0.0066 &  -0.0056 &   0.0008 &  -0.0164 &   0.0008\\
     26 May 2019 &   83.06 &     -0.0495 &   0.0012 &  -0.0089 &   0.0003 &  -0.0009 &   0.0031 &  -0.0051 &   0.0003 &  -0.0145 &   0.0003\\
     31 May 2019 &  129.83 &      0.1517 &   0.0048 &   0.0908 &   0.0008 &   0.1919 &   0.0040 &   0.0548 &   0.0012 &   0.0132 &   0.0012\\
     6 June 2019 &   92.47 &     -0.0337 &   0.0040 &  -0.0169 &   0.0015 &  -0.0166 &   0.0111 &   0.0080 &   0.0015 &  -0.0423 &   0.0014\\ 
    26 Sept. 2019 &    8.69 &     -0.1483 &   0.0005 &  -0.0323 &   0.0002 &  -0.1204 &   0.0002 &  -0.0186 &   0.0002 &  -0.0100 &   0.0001\\ 
      5 Oct. 2019 &   37.62 &     -0.1447 &   0.0011 &  -0.0054 &   0.0004 &  -0.0542 &   0.0014 &  -0.0030 &   0.0004 &   0.0082 &   0.0003\\ 
     12 Oct. 2019 &   38.41 &     -0.1198 &   0.0010 &  -0.0205 &   0.0003 &  -0.0654 &   0.0011 &  -0.0091 &   0.0003 &  -0.0043 &   0.0002\\ 
     19 Oct. 2019 &   41.59 &     -0.1441 &   0.0008 &  -0.0141 &   0.0003 &  -0.0622 &   0.0008 &  -0.0064 &   0.0003 &   0.0008 &   0.0002\\ 
     26 Oct. 2019 &   34.32 &     -0.1117 &   0.0001 &  -0.0232 &   0.0003 &  -0.0754 &   0.0012 &  -0.0106 &   0.0003 &  -0.0053 &   0.0002\\ 
     21 Jan. 2020 &   75.63 &     -0.0731 &   0.0012 &  -0.0013 &   0.0003 &  -0.0264 &   0.0016 &   0.0006 &   0.0004 &   0.0232 &   0.0003\\ 
     11 Feb. 2020 &   27.05 &     -0.0027 &   0.0019 &  -0.0418 &   0.0004 &  -0.0583 &   0.0007 &  -0.0197 &   0.0003 &   0.0089 &   0.0004\\ 
 \hline                                   
\end{tabular}
}
   \end{table*}

%

The OVIRS spectra show the ubiquitous presence of a 2.74 $\mu$m band feature due to hydrated minerals (Fig.~\ref{spectra}), which indicates that the parent body of Bennu experienced aqueous alteration (Hamilton et al. 2019). Although absorption features associated with hydrated minerals are detected on more than half of the C-complex main belt asteroids population (Fornasier et al. 2014; Howell et al. 2011; Rivkin et al. 2015), these  features are rarely detected on near-Earth asteroids. The 0.7 $\mu$m  band has been observed on the Ch asteroid (2099) Opik (Binzel et al. 2004), and potentially on (162173) Ryugu (Vilas, 2008), while the 2.7-3 $\mu$m band has been observed on (175706) 1996 FG3 (Rivkin et al. 2013) with a depth of 5--10 \%. In situ observations of the C-type Ryugu from the Hayabusa2 mission have shown a peculiar absorption centered at 2.72 $\mu$m, much weaker and narrower than that observed on Bennu, and consistent with thermally and/or shock-metamorphosed carbonaceous chondrite meteorites (Kitazato et al. 2019). \\
The fact that hydrated primitive asteroids are rare among the NEO population is probably related to the higher heating episodes undergone by their surfaces in the past, which diminished the depth of the hydrated absorption bands. \\
For Bennu, we find that the 2.74 $\mu$m absorption is relatively deep (consistent with Hamilton et al. 2019, and Simon et al. 2020a), with a depth of 15--17 \% for $\alpha < 30^{o}$ (Fig.~\ref{band2e7}). The center band position is consistent with the highly aqueous altered meteorites CM2 of petrologic type 2.1 and 2.2 (Hamilton et al. 2019; Rubin et al. 2007), and with the Pallas-like asteroid class, following the Takir et al. (2012) classification based on the $\sim$3 $\mu$m band shape and position.\\
As shown in Fig.~\ref{spectra}, the 2.74 $\mu$m band is observed in all the data acquired at different observing conditions and even at large phase angles. We estimate the 2.74 $\mu$m band depth using the spectral index (SPINDEX) calculation software (Kaplan et al. 2020). The SPINDEX software band depth calculations use the absorption band depth definition of Clark and Roush (1984). For this study, we define the 2.67 $\mu$m and 3.3 $\mu$m wavelengths as the left and right sides of the band, and 2.74 $\mu$m as its center. \\
We show in Fig.~\ref{band2e7} the 2.74 $\mu$m band depth for the different observing runs.  In the 0--90 $^{\circ}$ phase ranges, the 2.74 $\mu$m band depth slightly decreases with the phase angle; however, the variations are mostly within the error bars. No firm conclusions may be drawn for the variation of the 2.74 $\mu$m band depth versus the phase angle, except that it decreases at very high phase angles ($\alpha$ = 130$^{\circ}$), where it reaches a minimum depth of 10--12\%.

\section{Spectral phase reddening}

\subsection{Analysis of the global surface of Bennu}

For the investigation of phase reddening, we limited our spectral slope analysis to wavelengths between 0.45 and 2.50 $\mu$m, avoiding longer wavelengths that may be affected by residuals on the thermal tail correction. \\
We analyzed the spectral slope in multiple wavelength ranges and computed the average value and associated error for each observing sequence (Table~\ref{meanslope}). The spectral phase reddening effect is small but evident in the globally averaged spectra acquired at different phase angles (Fig. ~\ref{spectra}) and in the histograms of the spectral slope distribution (Fig.~\ref{histo} shows histograms for the 0.55--2.5 $\mu$m slope).\\
The most striking evolution with respect to the phase angle is the change of the slope from slightly negative to positive values for phase angles larger than $\sim$100 $^o$. \\
We stress that the datasets acquired at high phase angles have lower ratios of signal to noise and are not optimized for mineralogical investigation. However, visual inspection of the individual spectra acquired at high phase angles and fulfilling the aforementioned selection criteria indicates that all have a positive spectral slope; this is confirmed in two different datasets from 3 and 31 May 2019.  \\
These high-phase-angle observations were also acquired at high incidence and emission angles, in the 57--75$^{o}$ range. Laboratory measurements on the CM2 meteorite Mukundpura have shown changes in the spectral slope when incidence, emission, and phase angles are high (Potin et al. 2019), with slopes values changing from 0.67 $\mu$m$^{-1}$ at phase angle 10$^{o}$ to 1.2 $\mu$m$^{-1}$ at phase angle 130$^{o}$. Therefore, the Bennu spectra at high phase angle are consistent with these laboratory experiments.   \\
 Moreover, pre-encounter spectroscopic studies of Bennu indicate some variability in the spectral slope from unresolved spectroscopy. For example, observations from the Infrared Telescope Facility (IRTF) acquired at a high phase angle (98$^o$) have a positive spectral slope of 0.03 $\mu$m$^{-1}$ (Davies et al. 2007). Other work found slightly positive spectral slopes (Binzel et al. 2015), although not related to high phase angles (observations were carried out at 13 $< \alpha < 55 ^ {o}$). These data corresponded to observations across the equatorial ridge of Bennu, which is spectrally redder (Fig.~\ref{maps2a}) than the rest of the surface of Bennu. From the observed spectral variability, Binzel et al. (2015) concluded that the equatorial ridge of Bennu might be dominated by fine-particulate material. This conclusion was not confirmed by the OSIRIS-REx observations, which instead revealed a large concentration of boulders in the equatorial region, with little apparent fine-grained regolith (Lauretta et al. 2019). 
 
\begin{figure}
   \centering
   \includegraphics[width=0.5\textwidth,angle=0]{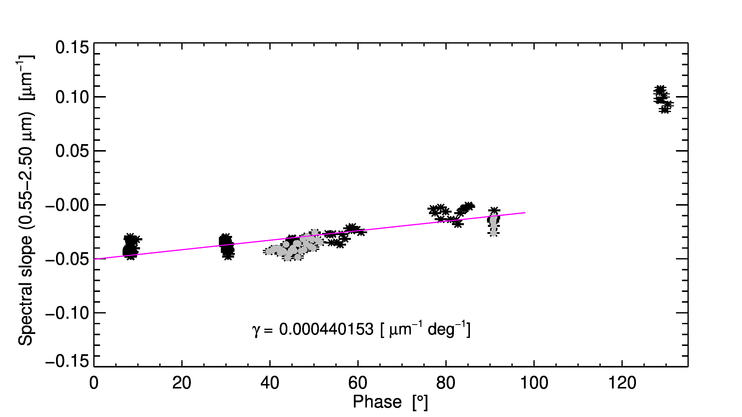}
   \caption{Phase reddening in spectral slopes evaluated in the 0.55--2.50 $\mu$m range. The black asterisks represent data acquired in the Detailed Survey from March to September 2019, while the gray circles those acquired during the Preliminary Survey in 2018. The magenta line indicates the linear fit of the Detailed Survey data, which have a higher resolution than the PS (see Table~\ref{observations}), for a phase angle lower than 100$^{\circ}$. Errors are smaller than the symbol size.}
              \label{reddeningall}%
    \end{figure}

\begin{figure}
   \centering
   \includegraphics[width=0.5\textwidth,angle=0]{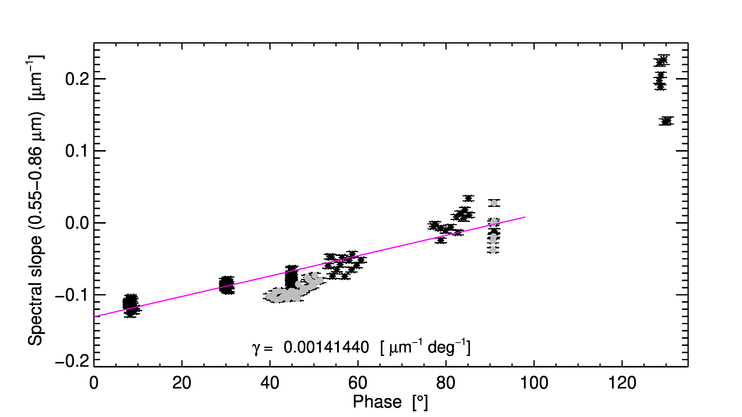}
   \caption{Phase reddening in spectral slopes evaluated in the 0.55--0.86 $\mu$m range. The black asterisks represent data acquired in the Detailed Survey from March to September 2019, while the gray circles those acquired during the Preliminary Survey in 2018. The magenta line is the linear fit of the Detailed Survey data, which have a higher resolution than the PS (see Table~\ref{observations}), for a phase angle lower than 100$^{\circ}$. Errors are smaller than the symbol size.}
              \label{reddeningvis}%
    \end{figure}

In our analysis of OVIRS data, we look in particular at two spectral slopes: 0.55--2.5 $\mu$m ($S2$ in Table~\ref{meanslope}; VIS and NIR), and 0.55--0.86 $\mu$m ($S3$; VIS). We chose these two wavelength ranges to facilitate the comparison with published studies on phase reddening effects of primordial bodies, which mostly cover the aforementioned ranges. The $S3$ slope is often noisier than that of $S2$ as it is computed across a narrower wavelength range that is covered by two OVIRS spectral segments (overlapping at $\sim$ 0.66 $\mu$m), which sometimes show discontinuities. \\
We show in Figures~\ref{reddeningall} and ~\ref{reddeningvis} the spectral slope in these two  wavelength ranges versus the phase angle for the data acquired during the Preliminary and Detailed Surveys. The phase reddening effect is linear up to $\alpha = 90^o$, but then it increases rapidly for higher phase angles. Similar nonlinear behavior with an exponential increase at high phase angles was observed for Mukundpura (Potin et al. 2019) in observations at high incidence and emission angles. \\
To compare the Bennu phase reddening values with published data, we computed the phase reddening coefficient using a linear fit of the Bennu slopes for $\alpha < 100^o$, where the slope variation is linear. Moreover, spectral data in the literature are usually available for phase angle $<$ 70--80$^o$. To avoid mixing different spatial resolutions, the linear fit (Figs.~\ref{reddeningall} and ~\ref{reddeningvis}) was evaluated using the data acquired from March to September 2019, which have a similar spatial resolution of 15--20 m. 
The resulting phase reddening coefficients and the estimated spectral slope at phase zero from the linear fit of the data are summarized in Table~\ref{gamma}.  \\
Bennu exhibits a moderate phase reddening effect with a monotonic phase dependence of the spectral slope; that is, the spectral slope continuously increases with increasing phase angle. Moreover, the phase reddening is wavelength-dependent: it is more prominent in the visible range and progressively decreases at longer wavelengths. In the 1.7--2.5 $\mu$m wavelength range, the  slope is almost constant for different phase angles, with a slightly negative phase reddening indicating a weak phase bluing at these wavelengths. The wavelength-dependent phase reddening of Bennu was also reported from OCAMS observations in the visible range by DellaGiustina et al. (2019). \\

The phase reddening phenomenon is generally attributed to multiple scattering in the surface medium at high phase angles and/or to small-scale surface roughness. Li et al. (2019) suggested that for low-albedo surfaces, the wavelength-dependent single scattering might be produced by micron-scale surface roughness or by small particles ($\sim$ micron sized).  This hypothesis is reinforced by laboratory experiments, which demonstrated that the micron-size particles and  the surface structure of larger grains, having roughness down to micron to sub-micron scales, give rise to a wavelength-dependent phase response, which produces phase reddening (Schr\"oder et al. 2014; Pilorget et al. 2016). 
The overall spectral phase reddening behavior of Bennu may thus be attributed to the presence of particles of microns to tens of microns, with fractal structure hosting micron to sub-micron roughness; that is, structures that have sizes comparable to those of the wavelengths measured by OVIRS.


 \begin{table}
      \caption[]{Phase reddening coefficients evaluated for different wavelength ranges. The quantity $\gamma$ is the phase reddening coefficient; Y$_0$ is the estimated spectral slope at zero phase angle from the linear fit of the data.}
         \label{gamma}
\centering                       
\begin{tabular}{c c c}       
\hline\hline 
Wavelength  & $\gamma$ & Y$_0$ \\
 range ($\mu$m)       &   ($\mu$m$^{-1} ~deg^{-1}$)       & ($\mu$m$^{-1}$) \\ \hline
0.55--2.50     & 0.000440$\pm$0.000028         & -0.050$\pm$0.004 \\
0.45--0.55    &  0.001162$\pm$0.000096         & -0.120$\pm$0.005 \\
0.55--0.86    &  0.001414$\pm$0.000100         & -0.130$\pm$0.009 \\
0.86--1.06    &  0.000517$\pm$0.000040         & -0.169$\pm$0.003 \\
1.08--1.70     & 0.000378$\pm$0.000052         & -0.045$\pm$0.003 \\
1.70--2.50    & -0.000042$\pm$0.000063         & -0.013$\pm$0.005 \\ \hline
\hline                          
\end{tabular}
   \end{table}

\begin{figure*}
   \centering
   \includegraphics[width=0.75\textwidth,angle=0]{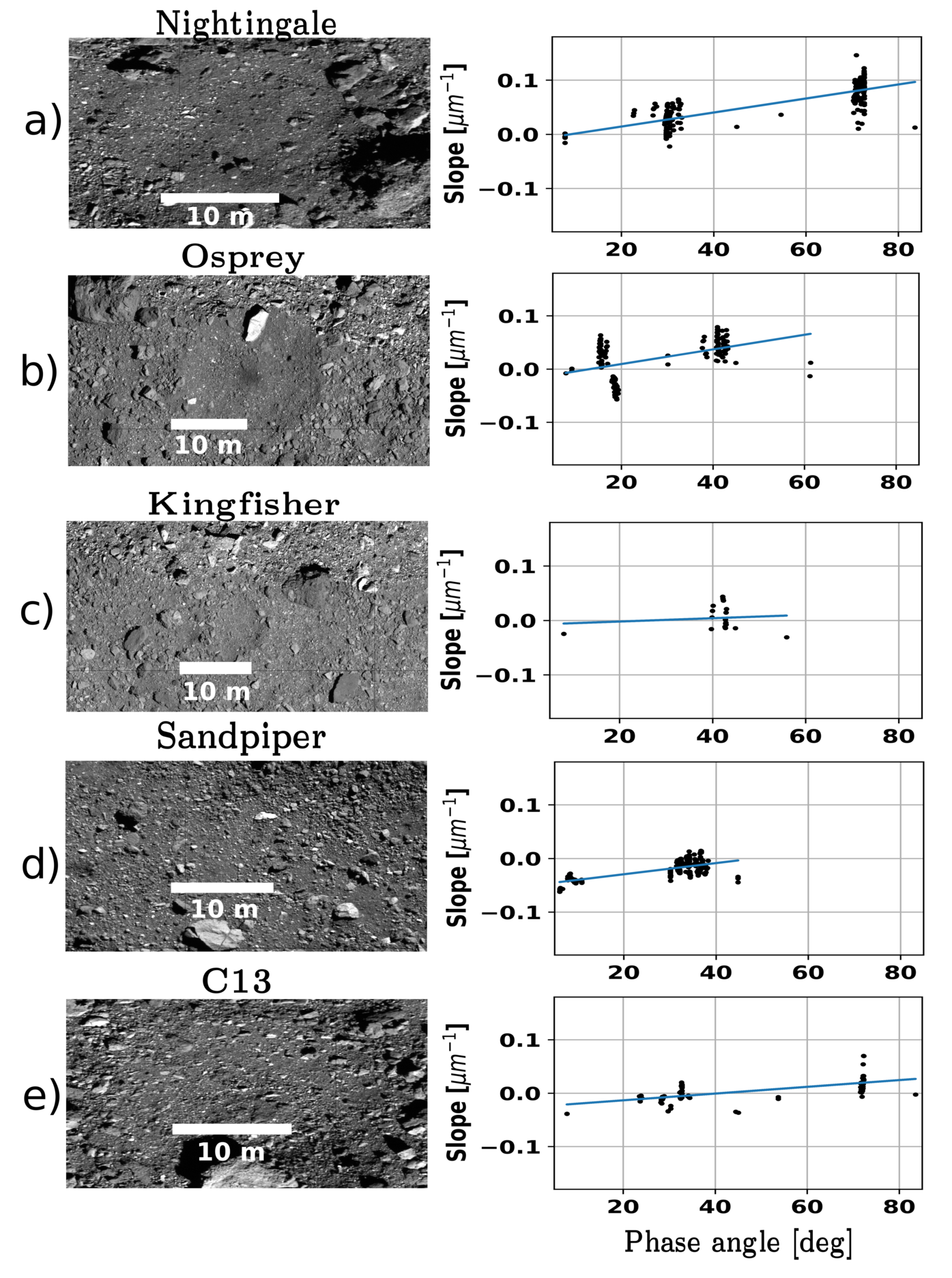}
   \caption{Images of the four candidate sampling sites and of the crater C13 (left) with the associated spectral phase reddening plots for the spectral slope evaluated in the 0.55--2.50 $\mu$m range (right). Coordinates are given in Table~\ref{gammaroi}. Images are taken from the Bennu global mosaic (Bennett et al. 2020). Each image is centered on a given ROI with a dimension of 10$^{o}$ in latitude (horizontal axis), and 20$^{o}$ in longitude (vertical axis).}
              \label{reddennightingale}%
    \end{figure*}

\begin{figure*}
   \centering
   \includegraphics[width=0.75\textwidth,angle=0]{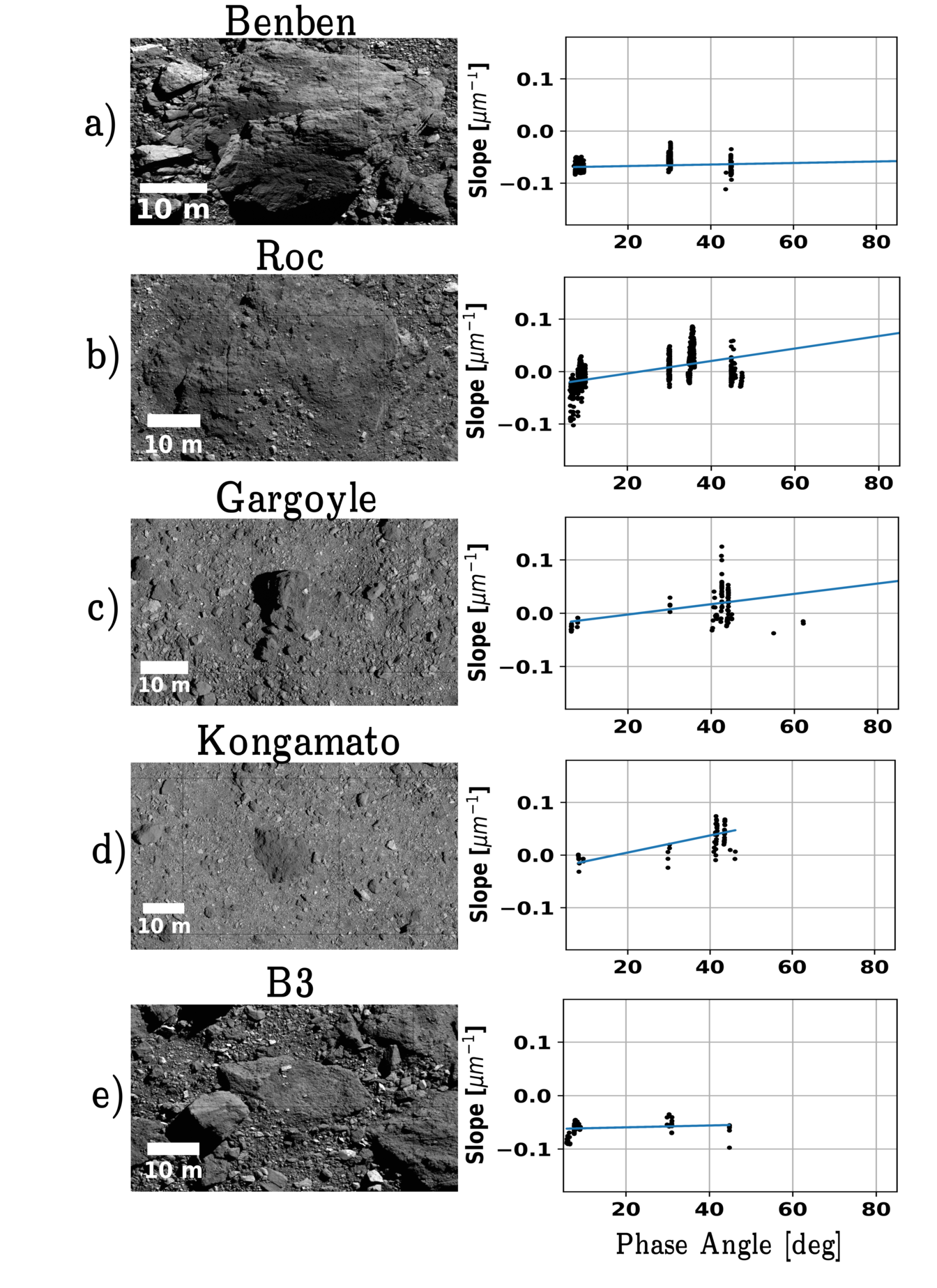}
   \caption{As in Fig.~\ref{reddennightingale}, but for five boulders. Coordinates are given in Table~\ref{gammaroi}. Images are taken from the Bennu global mosaic (Bennett et al. 2020).}
              \label{reddennigroi}%
    \end{figure*}

 \begin{table*}
      \caption[]{Coordinates and phase reddening parameters for the ROIs. The quantity $\gamma$ is the spectral reddening coefficient, and $Y$ the intercept at zero phase angle of the linear fit of the spectral slope vs. the phase angle. For craters (other than the four candidate sampling sites), the same ID numbers used by Deshapriya et al. (2020) are adopted, except for craters C46 to C49, which were not included in that 45-crater study and are arbitrarily numbered. For boulders, an arbitrary label (B1 to B5) is adopted, except for those that already have an official designation approved by the International Astronomical Union.}
         \label{gammaroi}
\centering
{\scriptsize                      
\begin{tabular}{c c c c c c c c c c c}       
\hline\hline 
         ROI    &     lat &     lon   & $\gamma_{0.55-2.5}$    & err$\gamma_{0.55-2.5}$ &          $Y_{0.55-2.5}$  &  err$Y_{0.55-2.5}$    &  $\gamma_{0.7-1.06}$ & err$\gamma_{0.7-1.06}$ & 
       $Y_{0.7-1.06}$  &  err$Y_{0.7-1.06}$  \\ 
                & ($^o$)  & ($^o$)  &     ($\mu$m$^{-1} ~deg^{-1}$) &  ($\mu$m$^{-1} ~deg^{-1}$) &    ($\mu$m$^{-1}$)  & ($\mu$m$^{-1}$) &  ($\mu$m$^{-1} ~deg^{-1}$) &  ($\mu$m$^{-1} ~deg^{-1}$)  &  ($\mu$m$^{-1}$)  & ($\mu$m$^{-1}$) \\  \hline
{\bf Craters}  & & & & & & & & & & \\ \hline
Kingfisher   &     11.0 &      56.0 &          0.000305 &              0.000627 &        -0.008182 &             0.026259 &          0.000619 &              0.000940 &        -0.073181 &             0.039380 \\
Nightingale  &     55.4 &      42.3 &          0.001294 &              0.000081 &       -0.011650 &             0.004600 &           0.0025063  &              0.000155 &        -0.1095088 &             0.008857 \\
Osprey       &     11.0 &      88.0 &          0.001296 &              0.000199 &        -0.014582 &             0.006229 &          0.000795 &              0.000177 &        -0.038713 &             0.005520 \\
Sandpiper    &    -47.0 &     322.0 &          0.000834 &              0.000055 &        -0.044621 &             0.001693 &          0.000958 &              0.000069 &        -0.121959 &             0.002122 \\
C1      &     -3.5 &     125.9 &         -0.000181 &              0.000058 &         0.005083 &             0.000986 &          0.000478 &              0.000066 &        -0.073673 &             0.001115 \\
C4      &     12.9 &     111.3 &          0.000524 &              0.000074 &        -0.053761 &             0.004003 &          0.000875 &              0.000070 &        -0.111653 &             0.003767 \\
C5      &     27.8 &      63.7 &         -0.000125 &              0.000025 &        -0.013623 &             0.000357 &          0.000346 &              0.000035 &        -0.098250 &             0.000503 \\
C6      &     -7.7 &     269.2 &          0.000294 &              0.000069 &        -0.009551 &             0.002530 &          0.000703 &              0.000105 &        -0.077026 &             0.003838 \\
C7      &    -43.3 &     325.2 &          0.001044 &              0.000045 &        -0.055259 &             0.001200 &          0.001355 &              0.000061 &        -0.140529 &             0.001652 \\
C8      &     -1.2 &     152.6 &          0.000081 &              0.000055 &        -0.034219 &             0.002028 &          0.000334 &              0.000077 &        -0.106002 &             0.002832 \\
C9      &      0.6 &     114.1 &          0.000152 &              0.000054 &        -0.035312 &             0.001614 &          0.000494 &              0.000080 &        -0.108977 &             0.002400 \\
C11     &      0.4 &     302.2 &          0.000127 &              0.000089 &        -0.028022 &             0.002868 &          0.000307 &              0.000148 &        -0.098102 &             0.004438 \\
C13     &     53.3 &      68.1 &          0.000536 &              0.000093 &        -0.023211 &             0.004573 &          0.000924 &              0.000179 &        -0.087708 &             0.008903 \\
C14     &      9.0 &      61.3 &          0.000218 &              0.000308 &        -0.024471 &             0.014922 &          0.000648 &              0.000472 &        -0.096850 &             0.022877 \\
C15     &     22.3 &      48.9 &          0.001744 &              0.000281 &        -0.060904 &             0.010172 &          0.002159 &              0.000373 &        -0.134144 &             0.013499 \\
C16     &     25.7 &      50.8 &          0.000674 &              0.000228 &        -0.028244 &             0.008956 &          0.001534 &              0.000366 &        -0.096334 &             0.014342 \\
C19     &     26.8 &     333.3 &          0.001694 &              0.000263 &        -0.026389 &             0.016965 &          0.002089 &              0.000830 &        -0.093405 &             0.052524 \\
C20     &     32.0 &     229.8 &          0.000639 &              0.000110 &        -0.041959 &             0.003579 &          0.000678 &              0.000218 &        -0.112865 &             0.005099 \\
C21     &     35.1 &     248.0 &          0.000539 &              0.000237 &         0.000182 &             0.009095 &          0.000806 &              0.000348 &        -0.062200 &             0.013362 \\
C22     &    -24.8 &     136.9 &          0.000250 &              0.000183 &        -0.020510 &             0.004045 &          0.000750 &              0.000165 &        -0.099391 &             0.003637 \\
C36     &     14.1 &     328.0 &          0.000905 &              0.000188 &        -0.059970 &             0.011888 &          0.001028 &              0.000373 &        -0.128348 &             0.023570 \\
C46         &     -4.0 &     125.0 &          0.000018 &              0.000053 &        -0.002986 &             0.001274 &          0.000381 &              0.000074 &        -0.077828 &             0.001777 \\ 
C47      &  -16.6    &     84.5     &    0.000109 &               0.000058 &           -0.035202 &             0.000547 &          0.000213 &              0.000095 &       -0.110507 &              0.000839 \\
C48         &      1.000 &     152.0 &          0.000107 &              0.000053 &        -0.029401 &             0.002030 &          0.000428 &              0.000077 &        -0.101266 &             0.002954 \\
C49       &    -53.8 &      65.0 &          0.000244 &              0.000071 &        -0.033335 &             0.000972 &          0.000396 &              0.000101 &        -0.113471 &             0.001387 \\
\hline
{\bf Boulders}  & & & & & & & & & & \\ \hline

Benben       &    -47.0 &     127.0 &          0.000094 &              0.000024 &        -0.069525 &             0.000593 &          0.000384 &              0.000048 &        -0.143367 &             0.001111 \\
Boobrie      &     49.0 &     216.0 &          0.000619 &              0.000090 &        -0.033773 &             0.003725 &          0.001200 &              0.000154 &        -0.112273 &             0.006727 \\
Camulatz     &      -10.0 &      260.0 &          0.0013576 &             0.000329 &        -0.003922 &             0.001095 &          0.001257 &              0.000507 &        -0.009482 &             0.001690 \\
Dodo         &    -34.0 &      64.0 &          0.000304 &              0.000112 &        -0.043333 &             0.003047 &          0.000234 &              0.000144 &        -0.106646 &             0.003734 \\
Gargoyle     &      5.0 &      93.0 &          0.000890 &              0.000217 &        -0.022002 &             0.008749 &          0.000202 &              0.000384 &        -0.076662 &             0.015453 \\
Gullikambi   &     19.0 &      19.0 &          0.000247 &              0.000047 &        -0.043963 &             0.002107 &          0.000565 &              0.000095 &        -0.117118 &             0.004392 \\
Hugin        &    -30.0 &      43.0 &          0.000975 &              0.000060 &        -0.031095 &             0.001752 &          0.001249 &              0.000084 &        -0.102925 &             0.002416 \\
Kongamato    &      5.0 &      67.0 &          0.001555 &              0.000255 &        -0.024983 &             0.010042 &          0.002241 &              0.000323 &        -0.085915 &             0.012790 \\
Roc          &    -24.0 &      28.0 &          0.000921 &              0.000064 &        -0.023989 &             0.001511 &          0.000493 &              0.000076 &        -0.063526 &             0.001759 \\
Tianuwa      &    -40.0 &     270.0 &          0.000686 &              0.000094 &        -0.058495 &             0.001208 &          0.000339 &              0.000137 &        -0.107947 &             0.001776 \\
B1  &     25.0 &     355.0 &          0.000213 &              0.000029 &        -0.042190 &             0.001202 &          0.000653 &              0.000048 &        -0.128416 &             0.001827 \\
B2        &    -18.0 &     256.0 &          0.000461 &              0.000066 &        -0.038616 &             0.000989 &          0.000600 &              0.000114 &        -0.089592 &             0.001787 \\
B3        &    -39.0 &     263.0 &          0.000175 &              0.000095 &        -0.060375 &             0.001704 &         -0.000155 &              0.000122 &        -0.117750 &             0.002275 \\
B4      &    -16.0 &     300.0 &          0.000076 &              0.000072 &        -0.015343 &             0.001012 &         -0.000033 &              0.000127 &        -0.066231 &             0.001773 \\
B5          &     14.5 &     262.0 &          0.000610 &              0.000133 &        -0.027675 &             0.004611 &          0.000468 &              0.000191 &        -0.085158 &             0.006657 \\
   \hline
     
\hline                                   
\end{tabular}
}
   \end{table*}
%
%
%
%

\begin{table*}
\caption{Spearman's rank correlation \emph{R} and $p-$value for craters and boulders between the phase reddening coefficient at multiple wavelengths and the phase function parameters.}
\label{pedro_spearman}
\noindent \begin{centering}
\begin{tabular}{|c|>{\centering}p{2cm}|>{\centering}p{2cm}|>{\centering}p{2cm}|>{\centering}p{2.5cm}|>{\centering}p{2cm}|>{\centering}p{2cm}|}
\hline 
                                      & ROI/craters & Linear albedo ($0.55\ \mu m$) & Phase function slope ($0.55\ \mu m$) & Spectral slope at $\alpha=0^{\circ}$ ($0.55-2.5\ \mu m$) & Linear albedo ($1.55\ \mu m$) & Phase function slope ($1.55\ \mu m$)\tabularnewline
\hline 
$\gamma_{(0.55-2.5)}$ & Craters   & 0.135, 0.53 & -0.211, 0.45 & -0.276, 0.19 & 0.10, 0.66 & 0.007, 0.97 \tabularnewline
\cline{2-7} 
                                       & Boulders  & 0.200, 0.48 & -0.087, 0.68 & 0.393, 0.147 & 0.157, 0.58 & -0.064, 0.82 \tabularnewline
\cline{2-7} 
$\gamma_{(0.7-1.06)}$  & Craters     & -0.030, 0.88 & 0.07, 0.75 & -0.235, 0.26 & -0.057, 0.79 & 0.150, 0.48 \tabularnewline
\cline{2-7} 
                                       & Boulders  & 0.125, 0.657 & -0.025, 0.9 & -0.121, 0.67 & 0.086, 0.76 & 0.157, 0.58 \tabularnewline
\cline{2-7} 
$\gamma_{(1.7-2.5)}$  & Craters     & -0.167, 0.44 & 0.243, 0.25 & 0.013, 0.95 & -0.146, 0.49 & 0.238, 0.26 \tabularnewline
\cline{2-7} 
                                       & Boulders  & 0.250, 0.37 & -0.350, 0.20 & 0.282, 0.31 & 0.218, 0.44 & -0.218, 0.435 \tabularnewline
\hline 
\multicolumn{1}{c}{} & \multicolumn{1}{>{\centering}p{2cm}}{} & \multicolumn{1}{>{\centering}p{2cm}}{} & \multicolumn{1}{>{\centering}p{2cm}}{} & \multicolumn{1}{>{\centering}p{2.5cm}}{} & \multicolumn{1}{>{\centering}p{2cm}}{} & \multicolumn{1}{>{\centering}p{2cm}}{}\tabularnewline
\end{tabular}
\par\end{centering}
\end{table*}

\subsection{Regions of interest}

Our study of phase reddening at local ROIs included the four candidate sampling sites, another 21 craters with well-defined rims (Bierhaus et al. 2020), and 15 prominent boulders (Table~\ref{gammaroi}). The boulders diameters range from 15 m to 95 m, and craters range from 5 m  to 160 m. Examples of spectral phase reddening plots for some ROIs are shown in Figs.~\ref{reddennightingale} and ~\ref{reddennigroi}.
The instrument spot size at the ground is about 15-20 m for most of the boulders and craters investigated, except for the four candidate sample collection sites, for which the spatial resolution on the ground was better by a factor of 4 to 5 (Table~\ref{observations}).

\subsubsection{Spectral phase reddening and morphological features}

In order to study whether morphological features and/or apparent surface texture are related to the phase reddening, we used the OCAMS global basemap of Bennu (Bennett et al. 2020) to examine ROIs with distinctive phase reddening properties.\\
We find that crater IDs 15 and 19 (C15 and C19) have the highest slope reddening coefficient($\gamma_{0.55-2.5}$) among all the investigated ROIs (Table~\ref{gammaroi}), almost four times higher than the Bennu global value. The C19 crater is 37 m in diameter with well-delimited rims and a smooth texture (at $\sim$0.06 m/pixel scale) with few meter-sized boulders. The C15 crater is smaller with a diameter of $\sim$ 9 m, but apparently rougher, with several boulders of 30 to 50 cm observed in the OCAMS global basemap. \\
Deshapriya et al. (2020) showed that C15 is the brightest and has the deepest 2.7 $\mu$m absorption band of the 45 craters they sampled, and this crater has a steeper spectral slope than the global Bennu average. The C19 crater has a higher slope and a slightly deeper 2.7 $\mu$m band than average, but its reflectance is very close to average. Deshapriya et al. (2020) suggested that C15 is younger than C19 and may have a composition that is less affected by space weathering processes. \\
The boulders Kongamato and Camulatz also show high phase reddening. Kongamato is darker than its surroundings with a rougher texture, while Camulatz  is a smoother boulder with inclusions on its upper face of 6 to 8 cm.  \\
Other outstanding features that have higher slope reddening than average by at least a factor 2 are the primary and back-up sampling sites of the OSIRIS-REx mission, Nightingale and Osprey,  contained in two craters that exhibit fine-particulate material, and the boulder Roc Saxum. These features have also steeper spectral slopes (Fig.~\ref{maps2a}). \\
On the basis of the coefficients listed in Table~\ref{gammaroi} and visualized in Fig.~\ref{pedro_one}, we do not observe a clear dichotomy among very different structures, such as craters and boulders, that look macroscopically smooth or rough. The fact that different morphological features share similar spectral phase reddening properties supports the hypothesis posited from the global data that phase reddening is driven by micro-roughness and particle fractal structures at very small scales.

\subsubsection{Spectral phase reddening and phase function parameters}

\begin{figure}[t]
   \centering
\includegraphics[width=0.45\textwidth,angle=0]{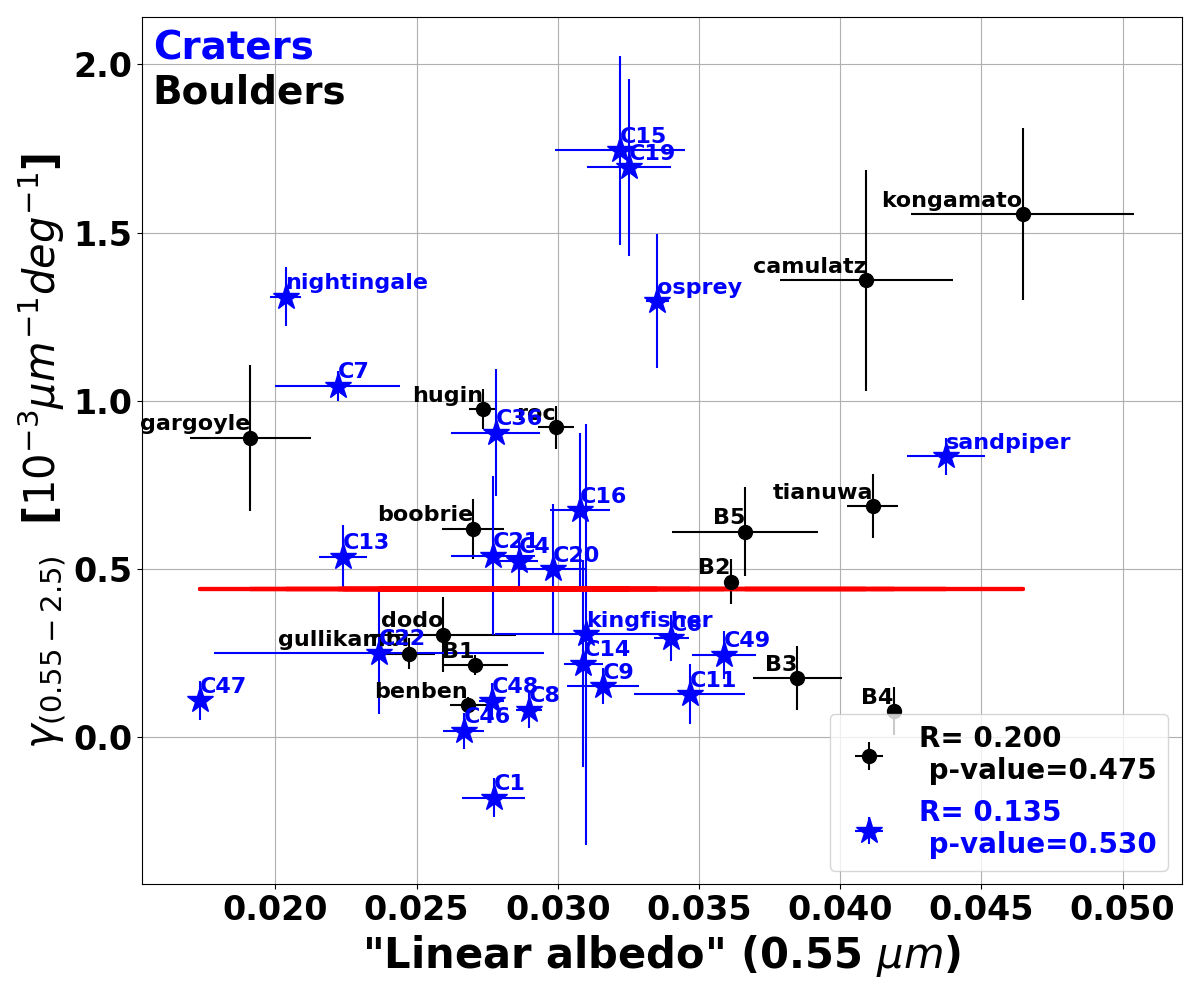}
\includegraphics[width=0.45\textwidth,angle=0]{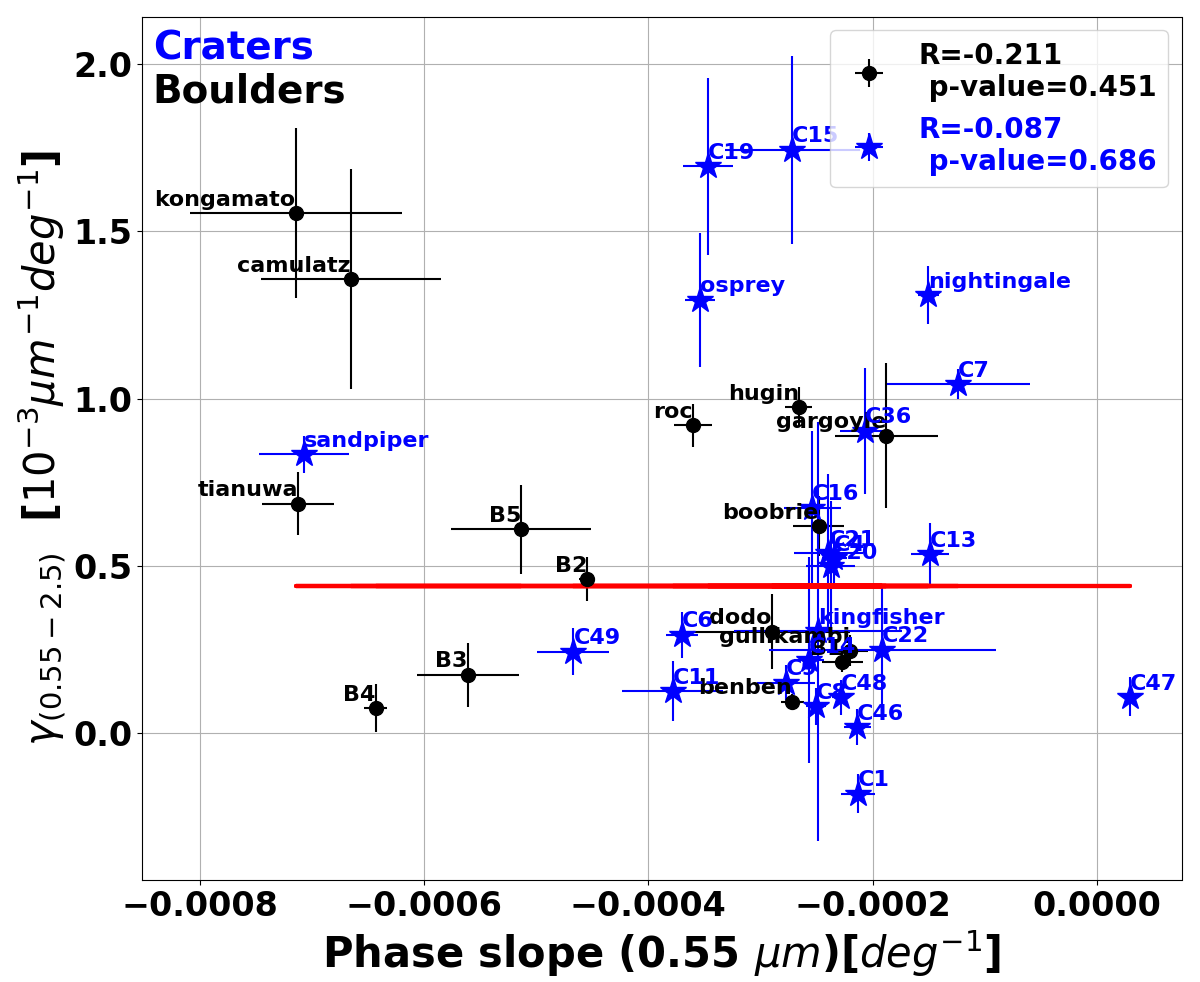}
\caption{Phase reddening $\gamma_{0.55-2.5}$  as a function of the linear albedo and of the linear phase function slope (both evaluated at wavelength $0.55\pm0.05\ \mu m$)
for the craters and boulders investigated. The red bar represents the average spectral phase reddening
of the surface of Bennu.}
\label{pedro_one}
\end{figure}

We investigated the correlations between phase reddening and phase
function parameters for craters and boulders separately. We computed the
Spearman's rank correlation $R$ and the $p-$value
for three spectral ranges and the linear albedo and phase
function slope at $0.55\ \mu m$ and at $1.55\ \mu m$. The \emph{R} and
$p-$values are listed in the Table~\ref{pedro_spearman}. The phase reddening and
phase function parameters are largely heteroscedastic; this is reflected
in the data scattering and the weak statistical significance of
the correlation coefficients.  However, boulders  generally show steeper phase function slopes than craters, as
shown in Fig.~\ref{pedro_one}. This behavior in unlikely to be associated with unresolved shadows, because they generally lead to lower signal-to-noise ratios or discontinuities in the spectra, and the data that have these kinds of artifacts were promptly remove from our sample. Boulders displaying the steepest phase function
slopes also have the highest linear albedos (see for example the boulders Kongamato,
Camulatz, Tianuwa, and B3--B5 in Fig.~\ref{pedro_one}).

Craters, on the
other hand, have most of their phase function slopes clustered between
$-4\cdot10^{-4}\ deg^{-1}$ and $-2\cdot10^{-4}\ deg^{-1}$, with
Sandpiper showing the steepest and highest phase function of all
studied craters. This different behavior between boulders and craters seems purely connected to the phase
function slope because there is no clear correlation between the spectral phase reddening
and the linear phase function parameters.\\
The apparent lack of correlations and the weak phase reddening coefficient observed may therefore be an indication that the roughness and/or particle size scales controlling the phase function and the spectral reddening are different.
Schr\"oder et al. (2014) suggest that roughness at scales smaller than the wavelength are needed to explain phase reddening. Conversely,  the phase function slope could be influenced primarily by size scales much larger than the wavelength.

\begin{figure}
\noindent \begin{raggedright}
\includegraphics[width=0.48\textwidth,angle=0]{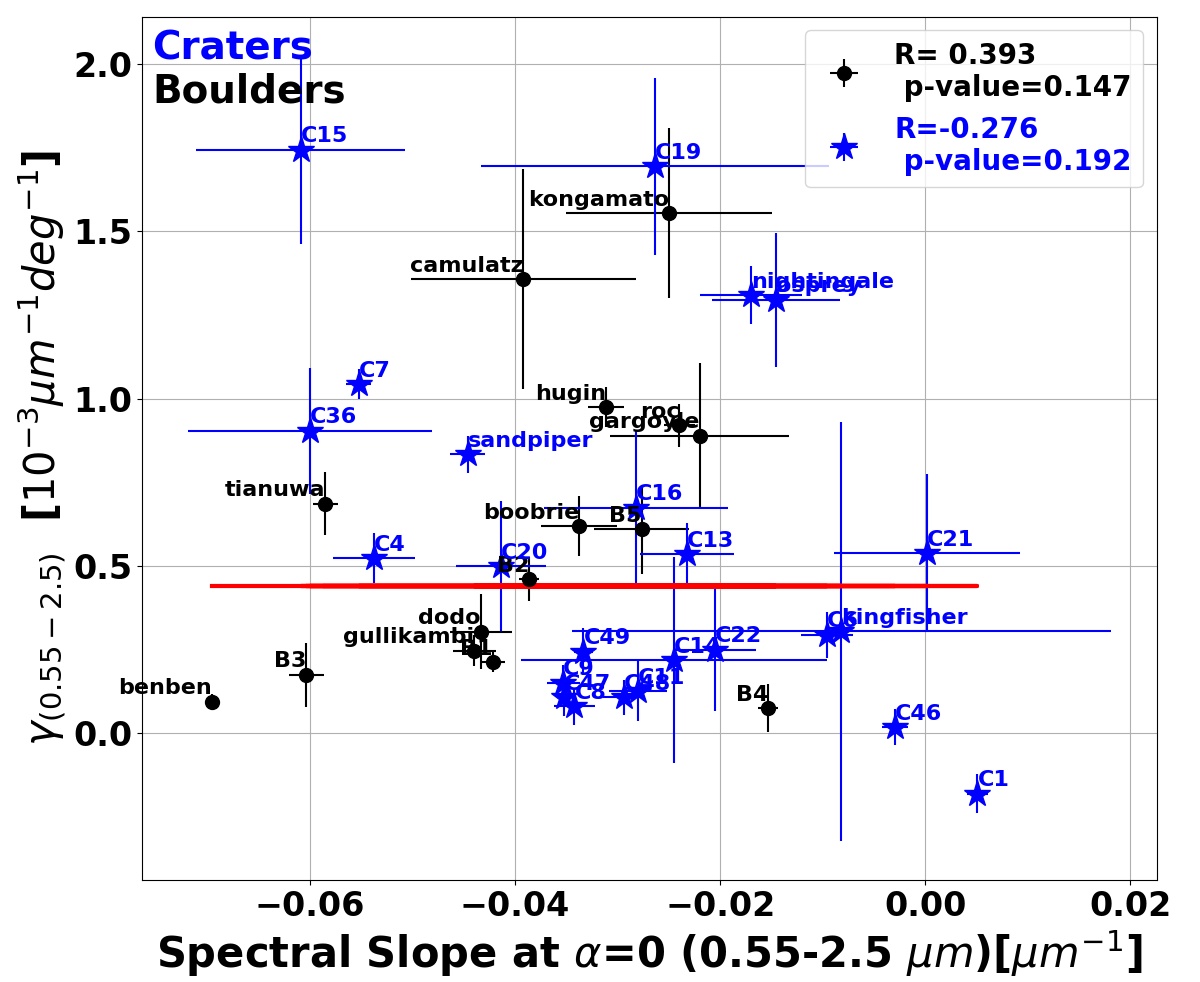}
\par\end{raggedright}
\caption{Phase reddening $\gamma_{(0.55-2.5)}$ as a function of the spectral
slope at $\alpha=0^{\circ}$ (derived from the 0.55--2.5 $\mu$m spectral slope) for craters (blue points) and boulders (black points). The red line represents
the average spectral phase reddening of the surface of Bennu.}
\label{pedro_bis}
\end{figure}

Nonetheless, there is a possible anticorrelation between the spectral slope at $\alpha=0^{\circ}$ and the spectral reddening
$\gamma_{(0.55-2.5)}$ among craters (Fig.~\ref{pedro_bis}, Table~\ref{pedro_spearman}), and a weak correlation among boulders,
indicating a difference between those terrains at very small scales. This anticorrelation is also found for the spectral reddening in the visible range ($\gamma_{(0.7-1.1)}$) for craters, but is weaker, and there is no significant anticorrelation at the longer wavelengths ($\gamma_{(1.7-2.5)}$). 
This wavelength-dependent behavior may be explained by the mechanism
proposed by Schr\"oder et al. (2014). 
Their numerical simulations suggest that phase reddening
is weaker in the NIR spectral range than in the UV-VIS one, as observed globally on Bennu (Table~\ref{gamma}).
In the same simulations, these authors produced a cluster of irregular particles of tens to hundreds of microns in size with smooth surfaces scattering specularly. By randomly adding Rayleigh diffusive scatterers to those particle surfaces, thus introducing surface roughness at micron and sub-micron scales, their simulations started mimicking the monotonic phase reddening observed on particulated opaque surfaces. The spectra were changed by reflectance dependence on $\lambda^{-4}$ for Rayleigh scatters. Therefore, overall, the mechanism is controlled by two scattering components, diffuse and specular. This is corroborated by the experiments on roughness of iron meteorites (Libourel et al. 2019), and possibly by results presented by Hasselmann et al. (2020). These authors showed that the  modelization of the bidirectional reflectance distribution of Bennu is improved when the two scattering components are both included in the model (Hasselmann et al. 2020).\\
Considering thus that the bluest spectral slopes are those more dominated by Rayleigh scatters on UV-VIS wavelengths, where the observed stronger phase reddening may be attributable to sub-micron roughness, the weak anticorrelation found for craters may indicate that they have a lower roughness size scale, possibly sub-micron sized, than boulders. On the other hand, the non-negative correlation observed for boulders suggests a larger roughness size scale, which may be on the order of microns.
 
\begin{figure}
\noindent \begin{raggedright}
\includegraphics[width=0.48\textwidth,angle=0]{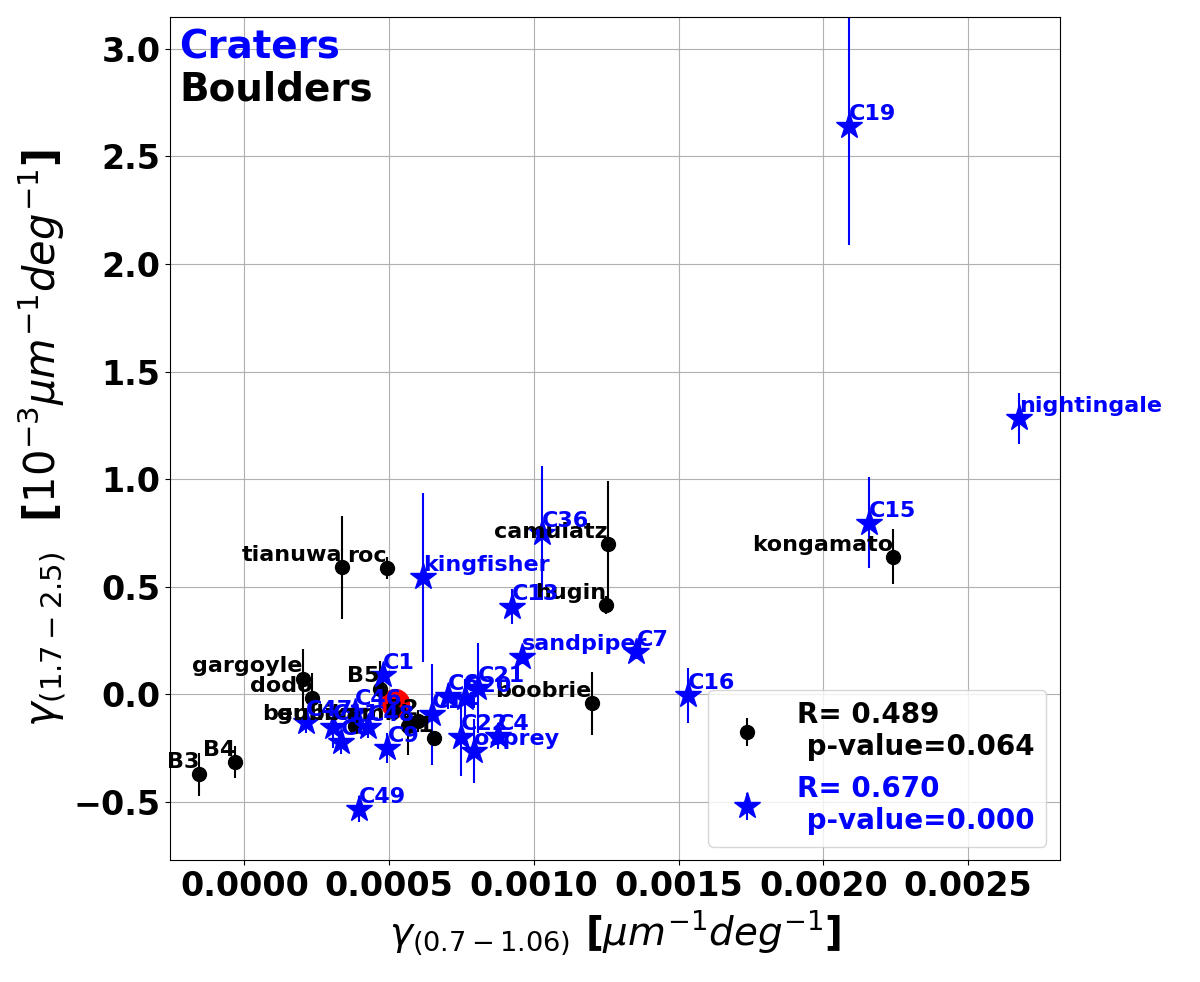}
\par\end{raggedright}
\caption{Phase reddening $\gamma_{1.7-2.5}$ vs. $\gamma_{0.7-1.06}$ for craters (blue points) and boulders (black points). The red circle represents
the average spectral phase reddening of whole Bennu's surface.}
\label{pedro_tris}
\end{figure}

We also find a significant correlation ($p-$value$<$14\%)
between the $\gamma_{(1.7-2.5)}$ and $\gamma_{(0.7-1.06)}$ for both boulders and craters (Fig.~\ref{pedro_tris}) that is strongest for craters. 
The results of Schr\"oder et al. (2014) may be compatible with phase reddening being an effect arising when the size scale of surface irregularities is comparable to the wavelength in which the phenomenon is measured. Correlations in phase reddening measured at various wavelength ranges may thus indicate that the characteristics of roughness are self-affine and scale invariant; that is, the particles roughness slope distribution is similar at different size scales, indicating the presence of fractal-like structures.  
The fact that phase reddening decreases at larger wavelengths (Table~\ref{pedro_spearman}) may provide an upper limit on the roughness size scale. Surface irregularities on Bennu seem more pronounced at $\sim$1 $\mu$m than at $\sim$2 $\mu$m size order from our phase reddening study, placing a possible constraint on their size scale in the VIS-NIR range.

\section{Discussion}

The variation of spectral slope with phase angle may be monotonic or arch-shaped.
Monotonic phase reddening effects have been observed on most of the Solar System bodies, with the notable exception of Mars (Guinness, 1981) and icy moons (Filacchione et al. 2012), which shows an arch-shaped change, first increasing (phase reddening) then decreasing (phase bluing). \\
Arch-shaped behavior is explained by multiple scattering and shadowing effects on surfaces dominated by semitransparent and relatively bright particles (Kaydash et al. 2010), while monotonic behavior is related to particle single scattering and/or wavelength-sized roughness on the particle surface, as shown by numerical simulations and laboratory experiments (Schr\"oder et al. 2014; Grynko and Shkuratov, 2008). These authors found that smooth surfaces give rise to arch-shaped reddening effects, while  microscopically rough regolith produces a monotonic phase reddening (Schr\"oder et al. 2014; Grynko and Shkuratov, 2008). In addition, Grynko and Shkuratov (2008) found that smooth particles larger than 250 $\mu$m produce phase bluing instead of reddening. \\

Our analysis indicates that the Bennu slope phase reddening effect is weak and wavelength-dependent, and affects the visible range more than longer NIR wavelengths. Very similar phase reddening coefficients and reddening behavior are also found by Li et al. (2020) and Zou et al. (2020) from the photometric modeling of OVIRS spectra. Moreover, DellaGiustina et al. (2019) and Golish et al. (2020) report a gentle phase reddening in the visible range of about 5\% at 90$^{o}$ phase for Bennu from analysis of OCAMS data.

\subsection{Comparison with other low-albedo bodies}

The phase reddening coefficient of Bennu in the 0.55--2.5 $\mu$m wavelength range (Table ~\ref{gamma}) is slightly higher than that  derived by Lantz et al. (2018) in a similar wavelength range from individual observations of 13 B-type asteroids at phase angles from 8 to 65$^o$; they reported a phase reddening coefficient of 0.00033$\pm$0.00086 $\mu$m$^{-1} ~deg^{-1}$. It should be noted that the phase reddening coefficient from Lantz et al. (2018) was not derived from the same object observed at different phase angles, but from different asteroids of the same taxonomic class observed at a given phase angle, and this gives rise to large associated errors. Their phase reddening coefficient for B-type asteroids may also be affected by other mechanisms related to different surface mineralogical abundances, particle sizes, or  amounts of regolith between the individual asteroids. Lantz et al. (2018) also reported moderate phase reddening for seven C-type asteroids at phase angles from 8 to 35$^o$ (0.00050$\pm$0.00163 $\mu$m$^{-1} ~deg^{-1}$), and a larger reddening effect for the D-type asteroids. \\ 
Perna et al. (2018) studied the spectral phase reddening effect on NEAs smaller than 600 m, again combining the spectral slopes of different asteroids observed at distinct phase angles.  They investigated the spectral slope in the 0.44--0.65 $\mu$m range for the main taxonomic classes and found that NEAs in the C-complex (which includes the B-types) display no or limited phase reddening.  

We may also compare Bennu with the low-albedo objects Ryugu and Ceres, which have been recently studied by the Hayabusa 2 and Dawn missions, and with comet 67P/Churyumov-Gerasimenko, observed by the Rosetta mission. Ryugu is a dark C-type NEA that shows evidence of aqueous alteration. Tatsumi et al. (2020) reported a phase reddening coefficient of (2.0$\pm$0.7) $\times 10^{-3} \mu$m$^{-1} deg^{-1}$ in the visible range, comparable to that derived in this work for Bennu (Table~\ref{gamma}).   \\ 
For the dwarf planet Ceres, Ciarniello et al. (2017, 2020) reported a monotonic spectral phase reddening throughout visible to NIR wavelengths, with $\gamma$ values of 0.0046 $\mu$m$^{-1} deg^{-1}$ in the 0.55-0.8 $\mu$m range and 0.0015 $\mu$m$^{-1} deg^{-1}$ in the 1.2-2 $\mu$m range. Thus phase reddening is progressively less sensitive to wavelength in the IR, that is, it gets smaller at longer wavelengths. They also concluded that phase reddening can be caused by sub-micron roughness, or sub-micron grains, qualitatively supporting a similar interpretation for the phase reddening behavior of Bennu.\\
Ceres has thus a steeper phase reddening effect than Bennu, with reddening coefficients that are about three times those of Bennu (Table~\ref{gamma}).  Li et al. (2019) confirmed the monotonic phase reddening effect on Ceres from their analysis of the photometric behavior at various wavelengths and attributed it to particle single scattering and/or small-scale roughness.   \\

Significant phase reddening effects were detected  for comet 67P/Churyumov-Gerasimenko (Fornasier et al. 2015; Ciarniello et al. 2015; Longobardo et al. 2017),  which has an albedo of 6\%, and red spectra (with visible spectral slope ranging from 1 to 2 $\mu$m$^{-1}$) matched by mixture of organic compounds and opaque minerals. The reddening $\gamma$ coefficient of this comet was 0.01 $\mu$m$^{-1} ~deg^{-1}$  in the 0.55--0.88 $\mu$m wavelength range (Fornasier et al. 2015) when the cometary activity was low (heliocentric distances $>$ 2.5 AU). This value decreased by a factor of 2 close to perihelion ($\gamma$ =0.0041 $\mu$m$^{-1} ~deg^{-1}$; Fornasier et al. 2016, 2017) when the higher cometary activity removed part of the 67P nucleus dust mantle, exposing  the  water ice--enriched subsurface layers. Therefore, the lower phase reddening effect close to perihelion may have indicated reduced surface micro-roughness as dust layers were lifted up by the cometary activity. In the IR region the spectral slope reddening was lower (0.0013--0.0018 $\mu$m$^{-1} ~deg^{-1}$ in the 1 to 2 $\mu$m range as reported in Ciarniello et al. (2015), and Longobardo et al. (2017)), indicating that comet 67P also has a wavelength-dependent phase reddening effect. Comparing the phase reddening coefficients in the same spectral range, the 67P phase reddening effect is roughly three to seven times that of Bennu.

Spectral phase reddening is also observed in meteorites. Beck et al. (2012) reported phase reddening in samples of carbonaceous chondrites, ordinary chondrites, howardite--eucrite--diogenite achondrites, and lunar meteorites.  
Binzel et al. (2015) and Cloutis et al. (2018) found that the samples of the CM2 meteorite Murchison with small particle sizes ($<$ 100 $\mu$m) are brighter and redder than the samples with larger particles, and that the packing density of the samples changes their spectral behavior. Spectral changes related to the porosity and particle size of the samples were also reported by Cloutis et al. (2011a, 2011b, 2018) for different CI and CM2 meteorites: they found that reflectance spectra of slabs are more blue-sloped (in the 1.8/0.6 $\mu$m reflectance ratio) and darker than powdered samples.\\ 
Binzel et al. (2015) and Cloutis et al. (2018) also noticed that phase angle, particularly at high emission angles, has a large effect on overall reflectance for the Murchinson meteorite, and a similar effect has been observed for other CI and CM2 meteorites (Potin et al. 2019; Cloutis et al. 2018). Potin et al. (2019) found a nonlinear phase reddening effect for the meteorite Mukundpura, with an exponential increase of the spectral slopes for $\alpha > 90^o$. In their study,  spectral reddening is more pronounced for the meteorite powder sample. These authors also found that the phase reddening of powdered sample is relatively insensitive to the incident angle, whereas it strongly increases at high phase angles for the raw meteorite sample. \\
With respect to the selected ROIs investigated on Bennu, we do not have enough observations or sufficient signal-to-noise ratio to investigate the phase reddening at high phase angles. However, our results indicate no obvious differences in spectral phase reddening between craters, which are macroscopically dominated by finer particles, and rougher structures such as boulders. Recent analysis of OTES thermal spectroscopy indicates the widespread presence of a thin layer (a few to 10s of $\mu$m thick) of fine particles ($<$ $\sim$ 65-100 $\mu$m), with some of the greater relative abundances identified in regions containing large, rough boulders (Hamilton et al. 2020). This supports our findings on globally similar phase reddening behavior for boulders and craters.  Models of Bennu thermal inertia also support the possible presence of a very thin layer of dust ($<$ 50 $\mu$m) on the whole surface (Rozitis et al. 2020), which would have a minimal effect on the apparent thermal inertia. \\
Constraints on the particles size inferred by OTES measurements (tens of micron) and by our investigation on slope phase reddening (sub-micron to a few micron) are different but not incompatible. In fact, OVIRS probes the uppermost surface of Bennu, where scattering properties are sensitive to very small particles comparable in size to that of the incoming light, while OTES is sensitive to larger particles and sounds the first subsurface. Hamilton et al. (2020) showed that a $\sim$ 5 $\mu$m-thick coating of dust, with dominantly particles size in the  5--50 $\mu$m range, result in a very small change in total emissivity spectral contrast. 
The presence of dust blanketing the surface of Bennu was also suggested from the analysis of DellaGiustina et al. (2019), who first reported the spectral phase reddening evident in OCAMS data. The possible presence of relatively fine regolith on the top of Bennu is also inferred by radiative transfer modeling of OVIRS data, as the best fit of Bennu spectral behavior is reached with intimate mixtures of CM2 heated meteorites  with relatively small particle sizes in the 5--15 $\mu$m range (Merlin et al. 2020).\\
These findings support the hypothesis that the process causing the slope phase reddening is controlled by small roughness scales, probably associated with fractal-like particle structures and/or a coating of fine particulates over the surface of Bennu. 
Moreover, high-resolution images acquired by OSIRIS-REx reveal porous and rough surface features up to the centimeter scale (Golish et al. 2020), thus pointing to complex roughness structures at different spatial scales. The sample returned from Bennu will provide essential information on particle size and structures at micron and sub-micron scales, well beyond the spatial scales reached until now from the OSIRIS-REx observations.     

\section{Conclusions}
In this work, we analyzed the spectral phase reddening effect on Bennu from the observations acquired with the OVIRS spectrometer onboard the OSIRIS-REx spacecraft.   We find the following:\\
\begin{itemize}
\item Consistent with previous findings, Bennu has a negative spectral slope (mean value of -0.043 $\mu$m$^{-1}$ at phase = 8 $^{\circ}$) typical of the B-type asteroids. This slope gently increases monotonically with increasing phase angle, showing a weak spectral phase reddening. The reddening coefficient is 0.00044 $\mu$m$^{-1} ~deg^{-1}$ in the 0.55-2.5 $\mu$m range, and it is stronger in the VIS range than in the NIR range. Based on evidence from numerical simulations and laboratory experiments, this implies that the surface of Bennu is covered by fine particles on the scale of microns and sub-microns and/or by particles with fractal structure that introduce micro-roughness.  
\item The spectral reddening is linear until phase $\sim$ 90$^{\circ}$, then it increases exponentially at high phases (130$^{\circ}$), where the spectral slope becomes positive. A similar behavior was observed for the CM2 meteorite Mukundpura (Potin et al. 2019). 
\item The spectral phase reddening on Bennu is comparable to that observed for B-type asteroids and to that reported for Ryugu from Hayabusa 2 observations (Tatsumi et al. 2020). It is however lower by a factor $>$ 3 than that of the dwarf planet Ceres and comet 67P/CG.
\item Spectral reddening does not differ significantly among craters and boulders analyzed at the local scale, possibly indicating that a coating of fine particles with micro-scale roughness covers the surface of Bennu.
\end{itemize}

The sample returned from Bennu will be fundamental to cast light on the regolith structure, composition, and photometric properties for low-albedo bodies. In particular it will provide the ground truth for understanding how macro- and microscale effects influence their spectrophotometric properties.

\begin{acknowledgements}
We are grateful to the entire OSIRIS-REx Team for making the encounter with Bennu possible. This material is based on work supported
by NASA under Contract NNM10AA11C issued through the New Frontiers Program. SF, MAB, PHH, AP, MF, and JDPD acknowledge funding support from CNES. 
PHH. acknowledges funding support by DIM ACAV+ program by the Region Ile de France. We thank Catherine Wolner for editorial help. We thank the referee, M. Ciarniello, for his comments and suggestions that helped improving this manuscript.
\end{acknowledgements}

%
%

%


%

%

\begin{appendix}

\section{Table}

\begin{table*}
\caption[]{Linear phase function coefficients for different ROIs. For craters (other than the four candidate sampling sites), the same ID numbers used by Deshapriya et al. 2020 are adopted, except for craters C46 to C49, which were not included in that 45-crater study and are arbitrarily numbered. For boulders, an arbitrary label (B1 to B5) is adopted, except for those having already an official designation approved by the International Astronomical Union. }
         \label{phasefunction}
{\scriptsize
\begin{tabular}{lrrrrrrrr}
{} &  slope550 &  err\_slope550 &  linearalb550 &  err\_linearalb550 &  slope1550 &  err\_slope1550 &  linearalb1550 &  err\_linearalb1550 \\ \hline

{\bf Craters} &        & & & & & & & \\ \hline
Kingfisher   &      -0.000249 &           0.000075 &      0.031004 &          0.003226 &       -0.000299 &        0.000057 &       0.032988 &           0.002458 \\
Nightingale  &      -0.000134 &           0.000010 &      0.020198 &          0.000535 &       -0.000118 &        0.000010 &       0.020047 &           0.000557 \\
Osprey       &      -0.000354 &           0.000013 &      0.033527 &          0.000412 &       -0.000317 &        0.000013 &       0.032681 &           0.000407 \\
Sandpiper    &      -0.000707 &           0.000040 &      0.043747 &          0.001375 &       -0.000484 &        0.000035 &       0.035426 &           0.001195 \\
C1      &      -0.000213 &           0.000015 &      0.027726 &          0.001114 &       -0.000199 &        0.000015 &       0.026542 &           0.001082 \\
C4      &      -0.000234 &           0.000011 &      0.028618 &          0.000670 &       -0.000212 &        0.000010 &       0.026786 &           0.000614 \\
C5      &       0.000173 &           0.000000 &      0.010241 &          0.000000 &        0.000269 &        0.000012 &       0.005198 &           0.000000 \\
C6      &      -0.000370 &           0.000014 &      0.034013 &          0.000621 &       -0.000353 &        0.000014 &       0.033093 &           0.000623 \\
C7      &      -0.000124 &           0.000064 &      0.022205 &          0.002224 &       -0.000014 &        0.000058 &       0.017604 &           0.002053 \\
C8      &      -0.000250 &           0.000009 &      0.028977 &          0.000462 &       -0.000234 &        0.000009 &       0.027355 &           0.000447 \\
C9      &      -0.000277 &           0.000026 &      0.031597 &          0.001277 &       -0.000250 &        0.000025 &       0.029445 &           0.001220 \\
C11     &      -0.000378 &           0.000045 &      0.034667 &          0.001966 &       -0.000351 &        0.000043 &       0.032722 &           0.001885 \\
C13     &      -0.000149 &           0.000017 &      0.022387 &          0.000857 &       -0.000131 &        0.000016 &       0.021374 &           0.000800 \\
C14     &      -0.000257 &           0.000013 &      0.030912 &          0.000679 &       -0.000249 &        0.000020 &       0.030067 &           0.001013 \\
C15     &      -0.000272 &           0.000060 &      0.032215 &          0.002304 &       -0.000183 &        0.000062 &       0.028537 &           0.002369 \\
C16     &      -0.000254 &           0.000025 &      0.030798 &          0.001063 &       -0.000251 &        0.000028 &       0.030771 &           0.001160 \\
C19     &      -0.000347 &           0.000022 &      0.032514 &          0.001493 &       -0.000338 &        0.000023 &       0.032443 &           0.001532 \\
C20     &      -0.000260 &           0.000026 &      0.030555 &          0.001185 &       -0.000232 &        0.000025 &       0.028542 &           0.001117 \\
C21     &      -0.000239 &           0.000031 &      0.027703 &          0.001492 &       -0.000257 &        0.000024 &       0.028858 &           0.001116 \\
C22     &      -0.000191 &           0.000101 &      0.023678 &          0.005850 &       -0.000177 &        0.000096 &       0.022426 &           0.005579 \\
C36     &      -0.000207 &           0.000023 &      0.027797 &          0.001588 &       -0.000182 &        0.000023 &       0.025642 &           0.001607 \\
C46        &      -0.000214 &           0.000012 &      0.026673 &          0.000721 &       -0.000198 &     0.000011 &       0.025501 &           0.000704 \\ 
C47       &       0.000029 &           0.000004 &      0.017322 &          0.000145 &        0.000015 &      0.000001 &       0.016919 &           0.000018 \\
C48        &      -0.000228 &           0.000008 &      0.027660 &          0.000434 &       -0.000213 &     0.000008 &       0.026259 &           0.000429 \\
C49      &      -0.000467 &           0.000032 &      0.035887 &          0.001120 &       -0.000439 &       0.000033 &       0.034185 &           0.001166 \\ \hline
{\bf Boulders} &        & & & & & & & \\ \hline
Benben Saxum      &      -0.000272 &           0.000010 &      0.026828 &     0.000651 &       -0.000244 &        0.000009 &       0.024337 &           0.000575 \\
Boobrie      &      -0.000248 &           0.000023 &      0.027002 &          0.001083 &       -0.000222 &        0.000022 &       0.025435 &           0.001027 \\
Camulatz     &      -0.000665 &           0.000080 &      0.040928 &          0.003071 &       -0.000593 &        0.000079 &       0.038000 &           0.003053 \\
Dodo         &      -0.000290 &           0.000070 &      0.025941 &          0.002589 &       -0.000258 &        0.000064 &       0.024001 &           0.002352 \\
Gargoyle Saxum    &      -0.000188 &           0.000046 &      0.019111 &          0.002147 &       -0.000152 &   0.000042 &       0.017439 &           0.001974 \\
Gullikambi   &      -0.000220 &           0.000015 &      0.024742 &          0.000927 &       -0.000200 &        0.000014 &       0.023014 &           0.000857 \\
Hugin        &      -0.000266 &           0.000012 &      0.027348 &          0.000490 &       -0.000253 &        0.000012 &       0.026582 &           0.000477 \\
Kongamato    &      -0.000714 &           0.000095 &      0.046464 &          0.003949 &       -0.000445 &        0.000075 &       0.035752 &           0.003121 \\
Roc Saxum      &      -0.000360 &           0.000017 &      0.029945 &          0.000630 &       -0.000329 &      0.000015 &       0.028791 &           0.000553 \\
Tianuwa      &      -0.000712 &           0.000033 &      0.041158 &          0.000905 &       -0.000636 &        0.000028 &       0.037826 &           0.000797 \\
B1           &      -0.000227 &           0.000018 &      0.027062 &          0.001176 &       -0.000204 &        0.000017 &       0.025143 &           0.001110 \\
B2        &      -0.000455 &           0.000007 &      0.036130 &          0.000159 &       -0.000417 &           0.000007 &       0.034329 &           0.000141 \\
B3       &      -0.000561 &           0.000045 &      0.038473 &          0.001576 &       -0.000504 &            0.000041 &       0.034878 &           0.001420 \\
B4     &      -0.000643 &           0.000010 &      0.041898 &          0.000183 &       -0.000600 &              0.000008 &       0.039983 &           0.000152 \\
B5        &      -0.000513 &           0.000062 &      0.036641 &          0.002584 &       -0.000449 &           0.000059 &       0.033671 &           0.002465 \\

\end{tabular}
}
\end{table*}

\end{appendix}

\end{document}